\documentclass[12pt, journal, draftclsnofoot, onecolumn]{IEEEtran}
\makeatletter
\def\ps@headings{%
\def\@oddhead{\mbox{}\scriptsize\rightmark \hfil \thepage}%
\def\@evenhead{\scriptsize\thepage \hfil \leftmark\mbox{}}%
\def\@oddfoot{}%
\def\@evenfoot{}}
\makeatother 
\usepackage[dvips]{epsfig}
\usepackage[dvips]{graphicx}
\usepackage{latexsym}
\usepackage{amssymb,balance}
\usepackage{graphicx}
\usepackage{subfigure}
\begin{document}
\title{Technical Report: Multi-Carrier Position-Based Packet Forwarding in Wireless Sensor Networks*}
\author{Ahmed Bader and Karim Abed-Meraim\\Telecom ParisTech, Paris, France\\\{bader,abed\}@telecom-paristech.fr\\ \and Mohamed-Slim Alouini\\KAUST, Thuwal, Mekkah Province, Saudi Arabia\\slim.alouini@kaust.edu.sa

\thanks{* This work was supported by Qatar National Research Fund (a member of Qatar Foundation).}

}

\maketitle

\begin{abstract}
Beaconless position-based forwarding protocols have recently evolved
as a promising solution for packet forwarding in wireless sensor
networks. However, as the node density grows, the overhead incurred
in the process of relay selection grows significantly. As such,
end-to-end performance in terms of energy and latency is adversely
impacted. With the motivation of developing a packet forwarding
mechanism that is tolerant to node density, an alternative
position-based protocol is proposed in this paper. In contrast to
existing beaconless protocols, the proposed protocol is designed
such that it eliminates the need for potential relays to undergo a
relay selection process. Rather, any eligible relay may decide to
forward the packet ahead, thus significantly reducing the underlying
overhead. The operation of the proposed protocol is empowered by
exploiting favorable features of orthogonal frequency division
multiplexing (OFDM) at the physical layer. The end-to-end
performance of the proposed protocol is evaluated against existing
beaconless position-based protocols analytically and as well by
means of simulations. The proposed protocol is demonstrated in this
paper to be more efficient. In particular, it is shown that for the
same amount of energy the proposed protocol transports one bit from
source to destination much quicker.
\end{abstract}

\maketitle

\section{Introduction}\label{sec:intro}
\indent Wireless sensor networks (WSNs) are being increasingly
considered for a multitude of monitoring and tracking applications
in the fields of process automation, environmental monitoring, and
tactical operations \cite{wsn.survey.Yicka}. WSNs are not limited to
terrestrial deployment scenarios but rather extend to submarine
environments \cite{underwater.WSN.Akyildiz}. The WSN umbrella is
also being expanded to fit more specific applications such as smart
utility networks
\cite{WSN.smartgrid.Gungor,WSN.smart_cities.Onworld} and multimedia
applications \cite{WMSN.survey.Akyildiz(Elsevier)}. The sincere
interest in WSNs coupled with the wide spectrum of possible
applications have created a tangible research surge across all
layers of the protocol stack over the past decade. Yet, most
research efforts tend to fine tune the balance between performance
and cost. The most important factors influencing the design of WSNs
are: energy, latency, complexity, and scalability
\cite{wsn.survey.Yicka,WSN.book.Akyildiz}.\\
\indent Packet forwarding is indeed one of the research issues which
directly impacts all of the aforementioned design factors.
Position-based forwarding protocols have emerged as some of the most
efficient packet delivery solutions for WSNs
\cite{wsn.survey.Yicka}. This is mainly due to the fact that nodes
can locally make their forwarding decisions using very limited
knowledge of the overall network topology. However, classical
position-based forwarding relies on periodical exchange of beacons
in the form of location updates between neighboring nodes.
Evidently, the beacon exchange process can often lead to a waste of
bandwidth \cite{beaconless.comparison.Sanchez}. To alleviate this
shortcoming, beaconless position-based forwarding techniques have
evolved as being more efficient. Two of the earliest such protocols
reported in literature are Geographic Random Forwarding (GeRaF)
\cite{GeRaF.latency.zorzi,GeRaF.multihop.zorzi} and Beaconless
Routing (BLR) \cite{BLR.Heissen}. In beaconless position-based
forwarding, potential relays undergo a distributed selection process
whereby the node with the most favorable attributes (e.g. closeness
to destination) shall eventually win the contention
\cite{beaconless.comparison.Sanchez}. The sender of a packet first
issues a request-to-send (RTS) message. Upon the reception of this
message, potential relays lying within the sender's coverage zone
enter into a time-based contention phase. Each potential relay
triggers a timer whose expiry depends on a certain cost function.
The first node to have its timer expire will transmit a
clear-to-send (CTS) message on the next time slot. However, since
time is slotted it is quite probable for collisions to occur. A
secondary collision resolution phase will be required in that
case.\\
\indent The ideas presented in
\cite{GeRaF.latency.zorzi,GeRaF.multihop.zorzi} and
\cite{BLR.Heissen} have been well-accepted within the research
community and have been furthered and/or adapted in
\cite{contention.based.Fussler,IGF.He,MACRO.Ferrara,cost.minimizing.Rossi,M-GeRaF.performance.Odorizzi}.
In Contention-Based Forwarding (CBF)
\cite{contention.based.Fussler}, response time to a RTS message is
rather calculated as function of the advancement towards the
destination. In \cite{IGF.He}, a protocol dubbed as Implicit
Geographical Forwarding (IGF) is proposed. IGF utilizes residual
energy and progress towards the destination as joint criteria for
relay selection. On the other hand, MACRO \cite{MACRO.Ferrara}
weighs the response time of potential relays with the progress that
can be made per unit power. Authors of \cite{cost.minimizing.Rossi}
propose a technique called Cost- and Collision-Minimizing Routing
(CCMR) whereby contending relays dynamically adjust their cost
metrics during the selection process. In
\cite{M-GeRaF.performance.Odorizzi}, GeRaF is modified such that it
serves wireless sensor networks with multiple sinks.\\
\indent Beaconless position-based forwarding offered valuable
improvements compared to their beacon-based predecessors. However,
they are plagued by an overhead which rapidly grows with node
density. As node density grows, collisions between candidate relays
contending for medium are more probable to occur, and thus the
overhead incurred grows noticeably. Not only does this degrade
end-to-end delay performance, but it also increases the mean energy
consumption. Furthermore, existing beaconless position-based
protocols have been built and evaluated based on the classical
``disc" coverage model. Under realistic multipath channel models,
this causes frequent duplication of packets and obviously leads to
the creation of co-channel interference
\cite{optim.beacons.Heissenbuttel}.\\
\indent In this paper, a novel position-based forwarding protocol is
proposed. The protocol's main virtue is that it does not resort to
any relay selection process. At any given hop, potential relays
check whether they satisfy certain position-based criteria. Any
relay that \emph{does satisfy} the criteria decides to forward the
packet ahead. It does so without reverting to any sort of
coordination with other potential relays. At the terminals of a
receiving node, such a mechanism would undoubtedly create multiple
copies of the same packet with different propagation delays. To
remedy this problem, we utilize a physical layer (PHY) which is
built over the use of orthogonal frequency division multiplexing
(OFDM). By virtue of introducing a cyclic prefix (CP) to each OFDM
symbol, a packet which is being simultaneously forwarded by multiple
nodes can be correctly detected at a receiving node
\cite{applications.OFDM}.\\
\indent Early generations of WSNs have actually employed low-power
Zigbee radios \cite{WSN.Zigbee.Baronti}. Zigbee is an IEEE
specification utilizing Direct Sequence Spread Spectrum (DSSS) and
is based on the IEEE 802.15.4 standard for wireless personal area
networks (WPANs)\cite{WSN.book.Akyildiz}. Nevertheless, considering
OFDM radios is a trend which has recently started to pick up and
gain traction \cite{SUN.PHY.overview,ofdm.balancing.Wu}. The
adoption of OFDM (also often referred to as multi-carrier
modulation) is mainly motivated by its ability to conveniently
accommodate larger channel bandwidths while featuring less
susceptibility to common radio channel impairments \cite{applications.OFDM}.\\
\indent The idea of using OFDM for concurrent transmissions of the
same packet has been actually proposed before in
\cite{sfn1.eriksson,sfn2.eriksson}. Nevertheless, the work of
\cite{sfn1.eriksson,sfn2.eriksson} was presented in the context of
utilizing OFDM for efficiently flooding a message across the whole
network. Obviously, this does not fit well into the WSN model where
data from sensors must be efficiently disseminated to a very limited
number of sinks. In the protocol presented in this paper, OFDM is
combined with position-based relaying in order to streamline packets
towards the intended destination.\\
\indent The protocol presented herein is the fruit of the marriage
between position-based relaying and OFDM. Nodes at a given hop $i$
will decide to relay or not based on the positions of the source,
the destination, and the relays of the previous hop, $i-1$. OFDM on
the other hand enables concurrent packet transmissions. The proposed
protocol dedicates part of the OFDM time-frequency resources for a
set of random access channel (RACH) slots. Each relay from the
$(i-1)$th hop randomly selects one of the RACH slots and modulates
it with its position information. When potential relays of the $i$th
hop receive the packet, they will read the position information on
all RACH slots and accordingly decide whether or not to relay.
Obviously, it is probable that some RACH slots may be selected by
more than one relay. Nonetheless, it will be demonstrated that such
``collisions'' on the RACH slots only affect the energy performance
as well as the achievable hop distances. In other words, they do not
impact the progression of the packet towards its destination.
Furthermore, it will be shown that the proposed protocol improves
the end-to-end delay performance when compared to existing
beaconless schemes. However, the amount of end-to-end energy
consumed is typically larger. This is due to the fact that one
packet is concurrently transmitted by multiple nodes every hop.
Interestingly, the improvement attained in terms of delay
performance is sufficiently large to offset the impact of additional
energy consumption. Indeed,
this is the major contribution of this paper.\\
\indent The rest of the paper is organized as follows. Section
\ref{sec:description} offers an elaborate description for the
operation of the protocol. Section \ref{section:PHY} presents a
model for the wireless channel and the physical layer. A detailed
statistical analysis of the hopping behavior under the proposed
protocol is given in section \ref{sec:hopping_dyn}. Evaluation of
the end-to-end performance against existing beaconless
position-based protocols and simulation results are provided in
section \ref{sec:eval}. Finally, the main results are summarized in
section \ref{sec:conclusions}.

\section{Description of Protocol Operation}\label{sec:description}
Since the proposed protocol is built around the concept of utilizing
OFDM, we will refer to it herein as OMR which is short of
``OFDM-based Multihop Relaying".

\subsection{Key Assumptions}
The following are key assumptions we used in the course of
developing OMR:
\begin{enumerate}
\item Positions of packet sinks are known to all nodes. Packet sinks may periodically advertise their own positions to the
rest of the network by means of a flooding process. In addition, all
nodes have knowledge of their own positions. In the analysis, the
source node $V$ is located in the cartesian space at $(0,0)$ while
the destination $Q$ is at $(L,0)$. \item Nodes are randomly
distributed in the network according to a 2-D Poisson point
distribution with density $\rho$. Furthermore, nodes implement a
non-synchronized sleeping schedule with a duty cycle of $\epsilon$.
\item If a node is awake but does not have data to transmit, then it
will be in a state of channel acquisition. In this state, the node
will be continuously monitoring the wireless medium for
synchronization pilots so as to be able to lock to any nearby
transmission / relaying process. \item A decode-and-forward relaying
strategy is adopted and nodes have omni-directional antennas.
\item Packet sinks are stationary. Nodes however can be fixed or
mobile. For packet durations in the range of milliseconds, the node
topology during one hop can be conveniently considered to stay
unchanged even for speeds of up to 100 km/hr.
\item A busy tone is activated during listening and receiving to
help mitigate the hidden node effect.
\end{enumerate}

\subsection{Packet Forwarding Process}
\begin{figure}[t]
\begin{center}
\epsfxsize=9cm \centerline{\epsffile{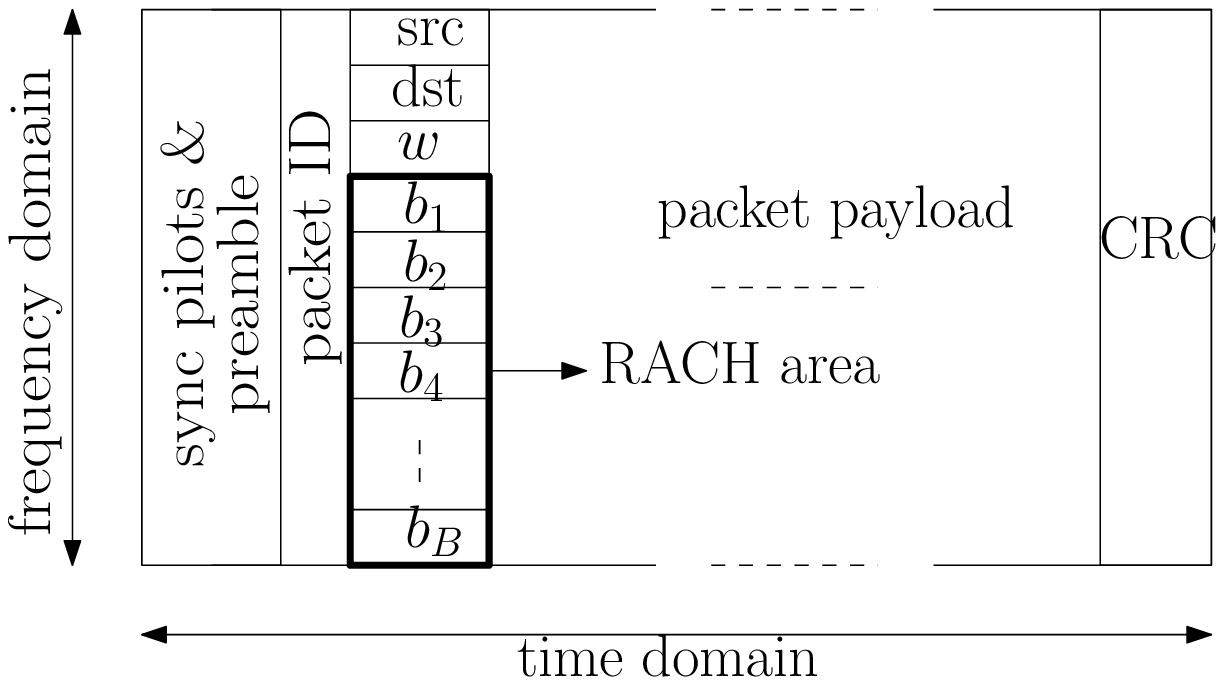}}
\end{center}
\caption{OMR packet structure.} \label{fig:packet_structure}
\end{figure}

\begin{figure}[t]
\begin{center}
\epsfxsize=12cm \centerline{\epsffile{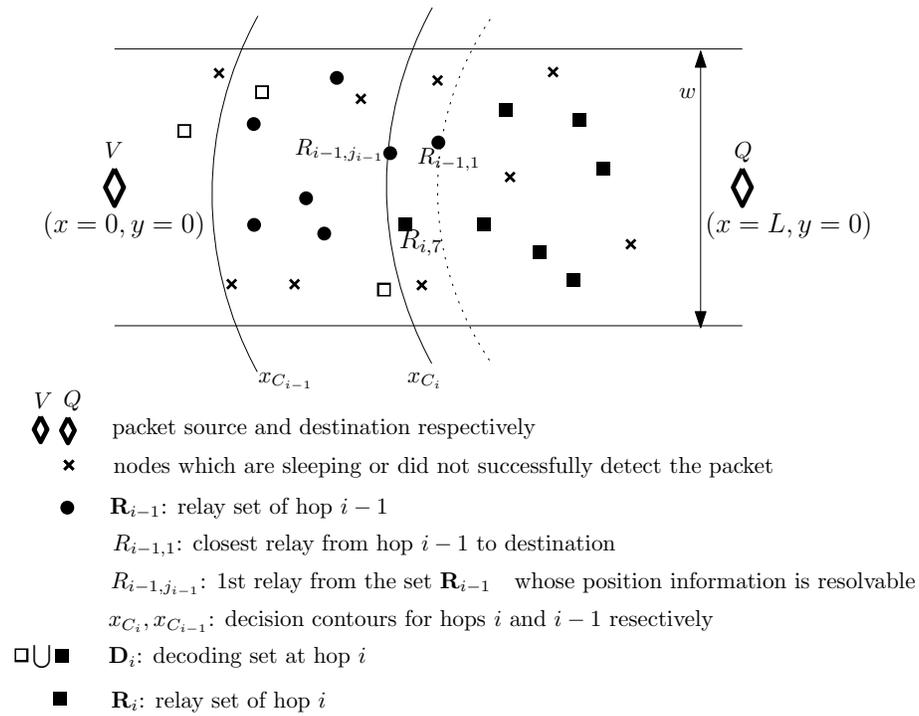}}
\end{center}
\caption{Example of a sample hopping process.}
\label{fig:typ_hopping}
\end{figure}

In the sequel, we will describe in detail how a packet is relayed
towards the destination. The source node $V$ will first encode its
own position information and that of the destination $Q$ on two
separate packet fields that are dedicated for this purpose. Figure
\ref{fig:packet_structure} illustrates the proposed packet structure
for OMR. Each packet is appended by a CRC sequence. The source node
then senses the wireless medium. The sensing activity must be
performed on the data tones as well as the busy tone
\cite{GeRaF.latency.zorzi,GeRaF.multihop.zorzi,dualBTMA.adhoc.Haas}.
If there exists any ongoing transmission in the vicinity it backs
off. Fixed window with binary exponential decrease back-off
algorithms may be employed to maintain proportional fairness in the
network \cite{wsn.survey.Akyildiz}. If the medium is not busy, the
source node will transmit the packet. Awaken nodes lying within the
transmission range of the source node will first attempt to decode
the packet and will activate the busy tone meanwhile. Nodes passing
the CRC check will become part of the decoding set of the first hop
denoted as $\mathbf{D}_1$. As a notational standard in this paper,
nodes in a given set are sorted in an ascending order according to
their distance to the destination $Q$. Two position-based criteria
are evaluated by each node in $\mathbf{D}_1$. The first is proximity
to the destination compared to the source. The second is whether it
lies within a forwarding strip of width $w$ surrounding the line
between $V$ and $Q$. The main objective of introducing the second
criterion is to streamline the forwarding process into a certain
geographical corridor. This will ensure that the interference
created by the forwarding process is confined to a certain
geographical region thus making more room for other forwarding
processes. Nodes in $\mathbf{D}_1$ satisfying the two position-based
criteria outlined above will become part of
the relaying set $\mathbf{R_{1}}$.\\
\indent To proceed, nodes in $\mathbf{R}_{1}$ will have to convey
their position information to nodes of the second hop. As shown in
figure \ref{fig:packet_structure}, the packet includes $B$ RACH
slots which are used for just that purpose. Each relay randomly
selects one of those RACH slots to modulate its position
information. Naturally, it is probable that one or more relays may
choose the same RACH slot. In such a case, it will not be possible
to resolve the position information of neither of those relays.
Nodes of the second hop are able to pick up the position information
of only those nodes in $\mathbf{R}_1$ which have selected unique
RACH slots. Relays of the set $\mathbf{R}_{1}$ will transmit the
packet after a small guard period. Awaken nodes which correctly
receive this transmission will now form the set $\mathbf{D}_2$. A
node in $\mathbf{D}_2$ will decide to forward the packet if: it is
inside the forwarding strip, and is closer to the destination than
all nodes of $\mathbf{R}_{1}$ whose position information could be
resolved. The forwarding process continues as described above for
the next hops and is further illustrated by an example in figure
\ref{fig:typ_hopping}.\\
\indent The source gives each packet a unique identification (ID)
label. Due to the sporadic nature of the wireless channel, a node
may receive the same packet a second time. By means of the unique
ID, the node will able to determine it is a duplicate packet and
thus will decide to drop it. This should happen before it continues
the decoding process, i.e. the node will not be part of the decoding
set of nodes.

\subsection{Effect of RACH Collisions}
The number of relays at hop $i$ is denoted by $\widetilde{K}_i$ such
that $\mathbf{R}_i=\{R_{i,k}\}_{k=1}^{\widetilde{K}_i}$. A RACH
collision is defined within this context as the event of having two
or more relays select the same RACH slot to modulate their position
information. As such, next-hop nodes will not be able to resolve the
positions of those relays. Recalling that $\mathbf{R}_i$ is an
ordered set, we define $(j_i)$ to be the index of the first node in
$\mathbf{R}_i$ whose position is resolvable, i.e. the positions of
the first $j_i-1$ nodes are unresolvable. Accordingly, it is
probable that some nodes at hop $i+1$ may relay the packet even
though they do not offer positive progress towards the destination.
In other words, they may be farther from the destination compared to
some or all of those $j_i-1$ relays. This is best illustrated by an
example. Looking at figure \ref{fig:typ_hopping}, node $R_{i,7}$
does not offer positive progress with respect to the node
$R_{i-1,1}\in\mathbf{R}_{i-1}$. Nonetheless, $R_{i,7}$ will actually
decide to relay since the position information of $R_{i-1,1}$
happens to be unresolvable. The first node in $\mathbf{R}_{i-1}$
whose position is resolvable is $R_{i-1,j_{i-1}}$, where in this
specific example $j_{i-1}=2$. The relaying criteria described above
can be formalized by first defining $x_{C_{i}}$ as the arc extended
from $Q$ and passing through node $R_{i-1,j_{i-1}}$. Subsequently, a
node in $\mathbf{D}_i$ will relay the packet if it lies in the area
$\between\{x_{C_{i}},y=\pm \frac{w}{2}\}$. Throughout this paper we
use the operator $\between$ to denote the area confined between
multiple contours.

\subsection{Carrier Sense Multiple Access at the Source}
We denote the number of nodes forming the decoding set at hop $i$ by
$\widetilde{L}_i=|\mathbf{D}_i|$ During hop $i$, the back-off region
produced by the forwarding process is composed of two subareas. The
first one is governed by the activity on the data channel, i.e. the
concurrent transmissions of $\widetilde{K}_{i-1}$ relays. The second
subarea is actually dictated by the superposition of the busy tone
signals from all $\widetilde{L}_i$ receiving nodes. As such, the
back-off region is substantially larger than the coverage zone of a
single node, e.g. a hidden node. Accordingly, relays in the case of
OMR can drop the task of sensing the medium before accessing it.
Channel sensing is only performed by the packet source node upon the
injection of a brand new packet. In contrast, relays in existing
beaconless protocols cannot afford as much not to listen to the
channel before transmitting. This is because the back-off region at
the final stages of the relay selection cycle is dictated by only
very few candidate relays. In other words, it will be only slightly
larger than the interference zone of a hidden node. Therefore, the
impact of a possible hidden node transmission is more drastically
felt in the case of traditional beaconless protocols. It is
worthwhile noting at this point that the hidden node issue has been
implicitly overlooked in
\cite{BLR.Heissen,contention.based.Fussler,IGF.He,MACRO.Ferrara,cost.minimizing.Rossi,M-GeRaF.performance.Odorizzi}.
On the other hand, it was ignored in \cite{GeRaF.latency.zorzi} and
\cite{GeRaF.multihop.zorzi} under the assumption of light to
moderate traffic loads.

\subsection{Retransmission Policy}
\begin{figure}[t]
\begin{center}
\subfigure[Expected number of retransmissions.]{
\includegraphics[width=13cm]{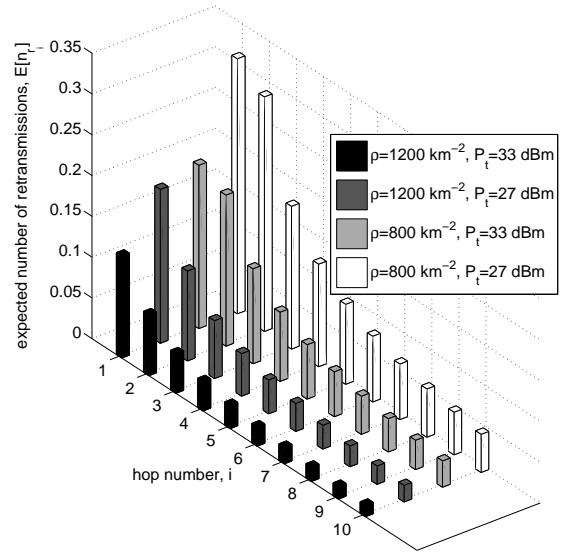}\label{fig:exp_retrans3}}
\subfigure[Rate at which false alarm events occur for various values
of the Packet ID decoding threshold. As more coding redundancy is
associated with the Packet ID field, a lower threshold will be
required.]{
\includegraphics[width=10cm]{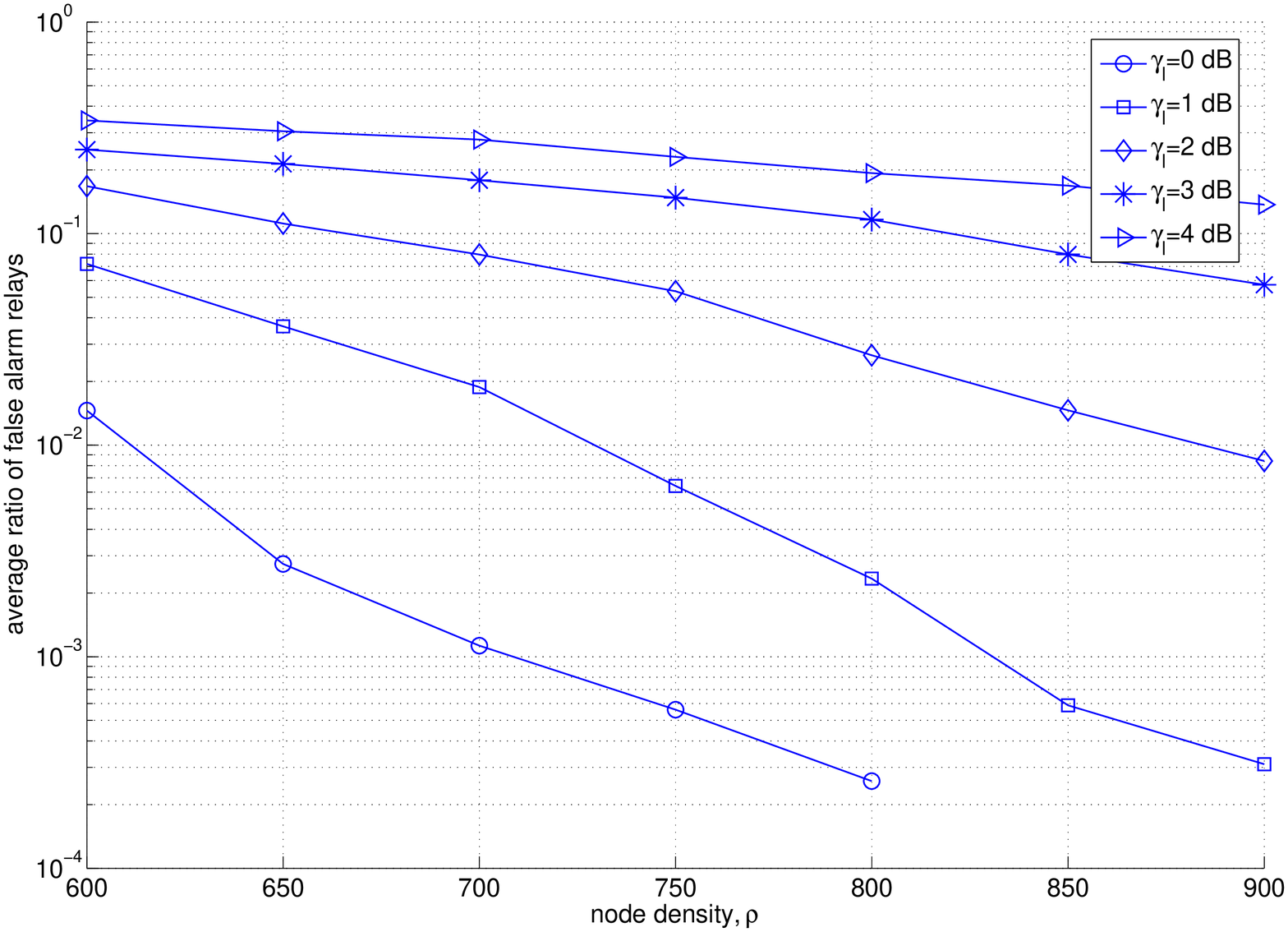}\label{fig:FA_occurrence_rate}}
\end{center}
\caption{Evaluation of the proposed retransmission policy.}
\label{fig:retransmission}
\end{figure}

\begin{figure}[t]
\begin{center}
\epsfxsize=10cm \centerline{\epsffile{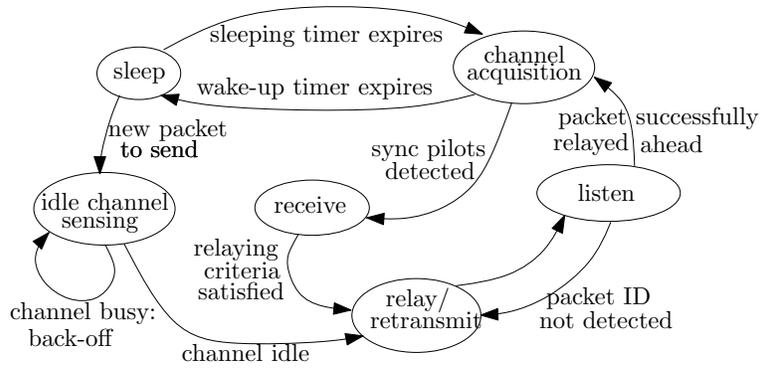}}
\end{center}
\caption{Simplified protocol state diagram.} \label{fig:state_diag}
\end{figure}

For the purpose of making OMR more reliable, a retransmission policy
must be devised. Nodes of $\mathbf{R}_i$ will need to retransmit if
the set $\mathbf{R}_{i+1}$ is empty, i.e. if there are no nodes
which decide to transmit the packet at hop $i+1$. To detect this
event, each relay in $\mathbf{R}_{i}$ must listen to the wireless
channel to verify that the packet is being forwarded ahead. A relay
in $\mathbf{R}_{i}$ will consider that the packet is being forwarded
if it is able to detect that packet's ID during the listening phase.
The packet ID is allocated a distinct field in the packet's header
as shown in figure \ref{fig:packet_structure}. For the sake of
saving energy, relays do not continue listening beyond the packet ID
time mark. Therefore, the listening activity only occurs for a
limited duration denoted here as $t_{ID}$. Powerful error coding
must be utilized for the packet ID field to make the listening task
more robust.\\
\indent Every time the packet is retransmitted, the width of the
forwarding strip, $w$, is increased by adjusting the corresponding
value in the packet header (figure \ref{fig:packet_structure}). This
reduces the probability that another retransmission would be
required. An expression for the expected number of retransmissions
at a given hop count is derived in section \ref{sec:hopping_dyn} and
is given in equation (\ref{eq:nri}). It is also plotted in figure
\ref{fig:exp_retrans3} for various values of node density and
transmit power. It is clear from the figure that the expected number
of retransmissions probability decreases as the packet progresses
towards the destination or as the node density increases. Similarly,
retransmission is less likely to occur as the transmit power
increases. Having described the retransmissions policy, we can now
summarize all different protocol states along with corresponding
state transitions in the diagram of figure \ref{fig:state_diag}.

\subsection{False Alarm Retransmissions}
A special case worthwhile investigating here is when a relay makes
an erroneous retransmission decision. This happens if a packet is
actually being forwarded ahead by nodes of $\mathbf{R}_{i+1}$, but a
relay from $\mathbf{R}_i$ believes otherwise, i.e. it was not able
to detect the packet ID during the listening phase. In order to
minimize the implications of a such a ``false alarm'' case, relays
must retransmit the packet only after the full packet duration
(denoted here by $T_p$) elapses. Moreover, the number of
retransmissions must be also capped to avoid infinite retransmission
loops by the false-alarm relay. If the number of retransmissions is
capped at $n_{r_{max}}$, then the impact of a false-alarm
retransmission event is only limited to incrementing the size of the
relay sets $\{\mathbf{R}_{i+n}\}_{n=2}^{n_{r_{max}+1}}$ by one. We
assume that the guard time of the OFDM symbol is in the range of
$120\mu s$ similar to \cite{SUN.PHY.overview}. Consequently, for a
false alarm retransmission to cause harmful interference, it has to
lie more than $7.2$ km away from the current hop. Typical dimensions
of a WSN makes us easily conclude that it is quite unlikely for
harmful interference to occur in the case of false-alarm
retransmissions. Furthermore, in case of a dense network and highly
redundant error coding of the packet ID, the false-alarm event
becomes even more unlikely to occur. Indeed, this intuition is
validated in figure \ref{fig:FA_occurrence_rate}. Accordingly, we
are practically able to neglect false-alarm retransmissions in our
subsequent analysis.

\subsection{Miscellaneous}
\indent Finally, it is worthwhile mentioning that we have evaluated
the potential of utilizing transmit diversity techniques for OMR. In
particular, we have considered the use of randomized transmit codes
\cite{cooperative.multihop.Scaglione} since they do not require
relays to coordinate their precoding matrices. However, the
implementation of such codes may prove to induce substantial
overhead as they mandate long training sequences for proper channel
estimation. Furthermore, they are suitable for the specific case of
narrow-band fading channels but not necessarily for wideband
channels.

\section{Physical Layer Modeling}\label{section:PHY}
In this section, a mathematical model for the overall channel
response is presented. Furthermore, a condition for successful
packet detection is developed.

\subsection{Wireless Channel Model}
\begin{figure}[t]
\begin{center}
\epsfxsize=9.5cm \centerline{\epsffile{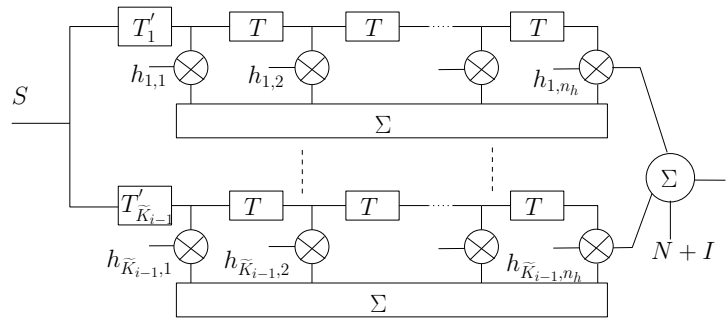}}
\end{center}
\caption{Wireless channel model.} \label{fig:ch_model}
\end{figure}

The channel between an arbitrary pair of nodes is represented by a
generic wideband multipath tap-delay line with Rayleigh-distributed
tap gains \cite{wireless.book.Rappaport} as shown in figure
\ref{fig:ch_model}. On average, there are $n_h$ such taps. Natural
echoes due to multipath are grouped in intervals of duration of $T$
seconds. The delays $T_1^{'}, \ldots,T_{\widetilde{K}_{i-1}}^{'}$
reflect the general case that the start of the $\widetilde{K}_{i-1}$
transmissions are not perfectly aligned in time. The duration of the
OFDM symbol is assumed to be larger than
$(n_h-1)T+\max\{T_k^{'}\}_{k=1}^{\widetilde{K}_{i-1}}-\min\{T_k^{'}\}_{k=1}^{\widetilde{K}_{i-1}}$
ensuring that each subcarrier encounters approximately a
frequency-flat fading \cite{applications.OFDM}. Amending each OFDM
symbols with a cyclic prefix eliminates inter-carrier interference
(ICI) and restores orthogonality between subcarriers. This enables
decoupled signal detection at each subcarrier. Given a certain
packet is relayed at hop $i$ by $\widetilde{K}_{i-1}$ nodes, then
the frequency response of the total channel at subcarrier $f_l$ is
given by $H(f_l)=\sum_{k=1}^{\widetilde{K}_{i-1}}e^{-j2\pi
f_lT_{k}^{'}}\sum_{n=1}^{n_h}h_{k,n}e^{-j2\pi f_l(n-1)T}$. It is
assumed that the duration of the cyclic prefix of the OFDM symbol is
long enough such that all signal echoes (natural and artificial)
arrive within the cyclic prefix interval. Other ongoing packet
relaying processes will rather contribute to the interference
signal. This interference however will be also Gaussian since the
individual channel gains are Gaussian \cite{hybrid.arq.Zhao}. The
exact nature of such an external interference is beyond the scope of
the present paper and is rather a subject of future work. Under the
reasonable assumption that the fading coefficients $h_{k,n}$ are all
mutually independent, it follows that $H(f_l)$ is complex Gaussian
such that $H(f_l)\sim\mathcal{N}(0,\sigma_S^2)$. $|H(f_l)|^2$ is
exponentially distributed with a mean of
$2\sigma_S^2=2\sum_{k=1}^{\widetilde{K}_{i-1}}\sum_{n=1}^{n_h}
\mathbb{E}[|h_{k,n}|^2]$. We note that $\sum_{n=1}^{n_h}
\mathbb{E}[|h_{k,n}|^2]$ represents the mean power content of the
channel between the receiver and the $k$th relay and is equal to
$\left(\lambda/4\pi\sqrt{(x-x_k)^2+(y-y_k)^2}\right)^\alpha$. Here
$\lambda$ represents the wavelength, $\alpha$ is the large-scale
path loss exponent, and $\sqrt{(x-x_k)^2+(y-y_k)^2}$ is the distance
from the receiver located at $(x,y)$ to the $k$th relay located at
$(x_k,y_k)$. Therefore, we obtain
\begin{equation}
\sigma_S^2=\left(\frac{\lambda}{4\pi}\right)^\alpha\sum_{k=1}^{\widetilde{K}_{i-1}}\frac{1}
{\left((x-x_k)^2+(y-y_k)^2\right)^\frac{\alpha}{2}}.\label{eq:sigma_S2}
\end{equation}
\indent Phase shift keying (PSK) modulation is utilized such that
the transmit power is equal for all subcarriers. The signal to
interference plus noise ratio (SINR) at subcarrier $f_l$ is given by
$\gamma(f_l)=\frac{P_t|H(f_l)|^2}{N_sP_n}$, where $P_t$ is the
transmit power over the whole bandwidth, $N_s$ is the number of
subcarriers, and $P_n$ is the noise plus interference power within
the subcarrier bandwidth. The average SINR is subcarrier-independent
and is denoted by $\gamma_o=\frac{2P_t}{N_sP_n} \sigma_S^2$.

\subsection{Successful Packet Detection}
The outage probability, $P_o$, is the probability that
$\gamma(f_l)<\gamma_t$ and equals $1-e^{-\gamma_t/\gamma_o}$. We
consider that a packet is successfully detected if $P_o<\tau$, where
$\tau$ is a detection reliability parameter
\cite{routing.fading.Haenggi}. The value of $\tau$ relies on the
underlying coding and interleaving techniques\footnote{With
reference to figure \ref{fig:packet_structure}, it is assumed that
interleaving is only applied to the payload and CRC portions of the
packet. This will enable relays in the listening phase to drop right
after the packet ID time mark.}. With $0<\tau<1$, and recalling that
$\gamma_o=\frac{2P_t}{N_sP_n} \sigma_S^2$, then the condition for
successful detection is expressed as
\begin{equation}
\sigma_S^2\geq \frac{N_sP_n\gamma_t}{2P_t
\ln\frac{1}{1-\tau}}\label{eq:sigma_ineq}.
\end{equation}
The mean coverage contour for hop $i$ is denoted by $x_{H_i}(y),
|y|\leq \frac{w}{2}$. Based on the successful packet detection
condition of (\ref{eq:sigma_ineq}) and in light of
(\ref{eq:sigma_S2}) we define
\begin{eqnarray}
x_{H_i}(y) & : & H_i(x_{H_i}(y),y)=U,\label{eq:xHi}
\end{eqnarray}
where
\begin{eqnarray}
H_i(x_{H_i}(y),y)&=&\sum_{k=1}^{\widetilde{K}_{i-1}}
\frac{1}{((x_{H_i}(y)-x_k)^2+(y-y_k)^2)^{\frac{\alpha}{2}}},\nonumber\\
U&=&\frac{N_sP_n}{2P_t}
\left(\frac{4\pi}{\lambda}\right)^\alpha\frac{\gamma_t}{\mathrm{ln}\frac{1}{1-\tau}}\nonumber.
\end{eqnarray}
For the first hop, we have
$x_{H_1}(y)=\sqrt{\frac{1}{U^{\frac{2}{\alpha}}}-y^2}$.

\section{Practical Considerations}\label{sec:prac_consider}

\subsection{Timing Considerations}\label{sec:timing}
\begin{figure}[t]
\begin{center}
\epsfxsize=10cm \centerline{\epsffile{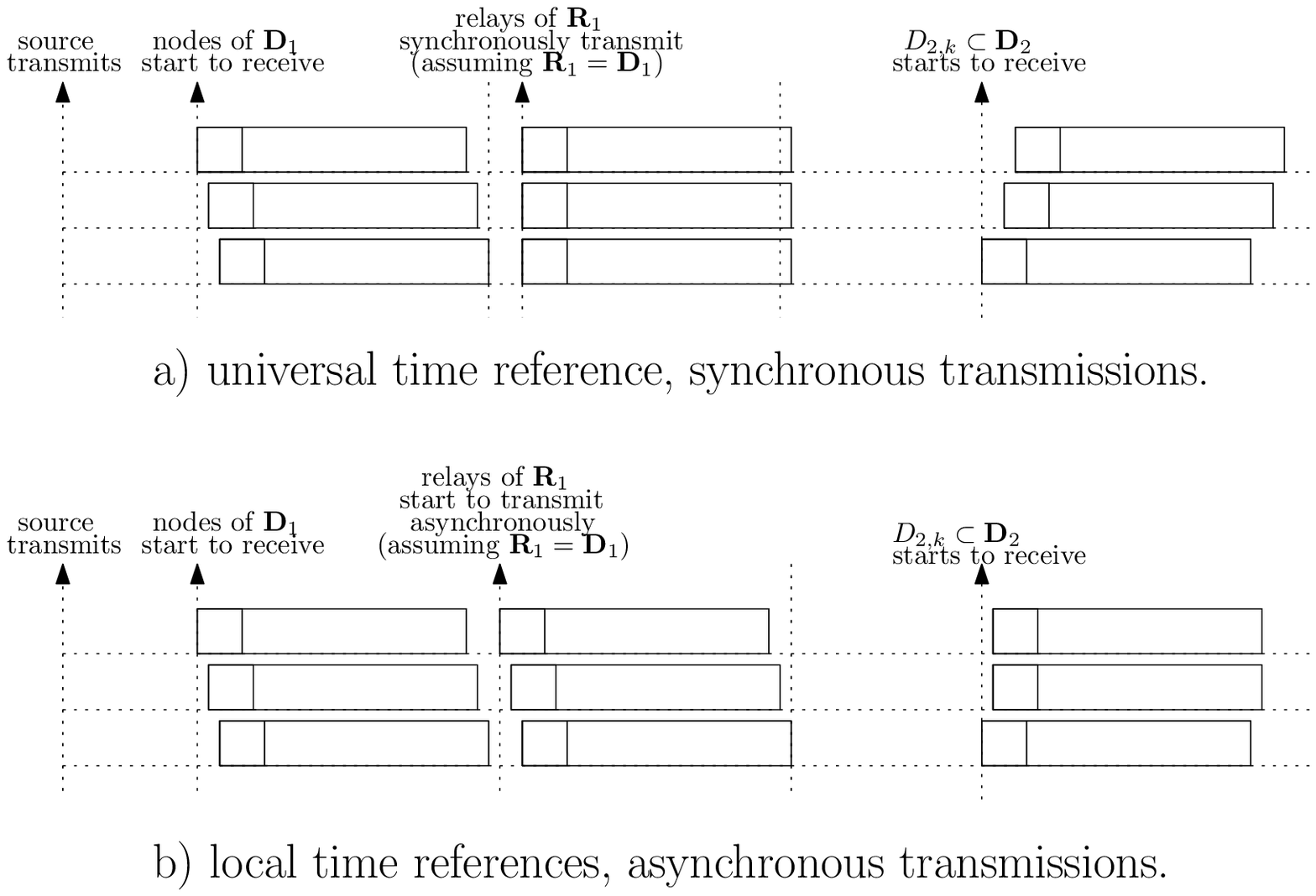}}
\end{center}
\caption{When universal time reference is available in the network
by means of GPS for example, synchronous relaying is possible at
each hop. A more general case however is to have asynchronous
transmissions.} \label{fig:time_reference_univ}
\end{figure}

\begin{figure}[t]
\begin{center}
\subfigure[Standard deviation and mean of the forwarding delay
spread as function of the strip width for various node densities
($\rho=900,1200,$ and $1500$ km$^{-2}$). The length of the strip is
600m.]{
\includegraphics[width=9cm]{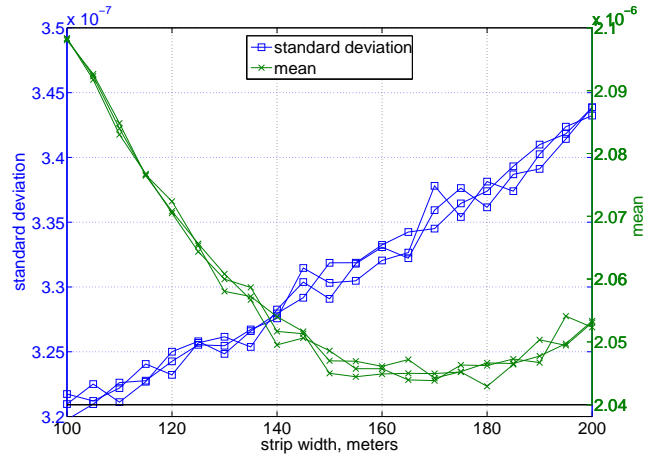}\label{fig:delay_spread_mean_std}}
\subfigure[Delay spread histogram ($\rho=1500$ km$^{-2}$, $w=200$m,
50000 iterations).]{
\includegraphics[width=9cm]{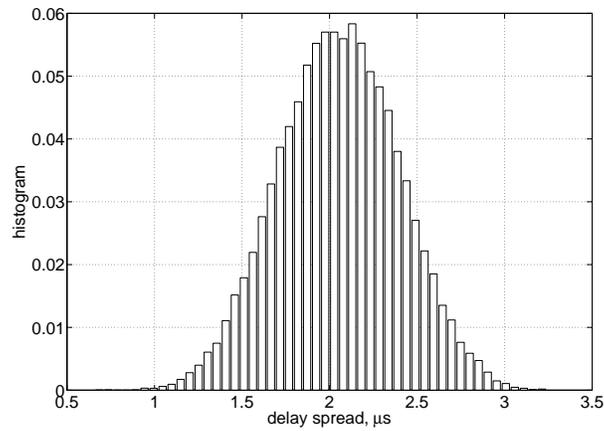}\label{fig:del_spr_hist}}
\end{center}
\caption{Statistics of the forwarding delay spread obtained through
simulations.} \label{fig:delay_spread}
\end{figure}

We recall our assumption that a fairly accurate clock and frequency
synchronization between nodes is attained via the same process by
which the position information is acquired \cite{wsn.survey.Yicka}.
In fact, the forwarding process itself may also aid to achieve the
same goal by means of the synchronization pilots inserted in the
packet header. Furthermore, if an external localization method such
as the global positioning system (GPS) is used, then nodes can be
also aligned to a universal time reference. However, GPS is known to
be power-hungry and thus is generally unfavorable for WSNs. When
other localization methods are utilized nevertheless, we must assume
that a universal time reference does not generally exist. Instead, a
node in the receiving state will align its time reference to the
first energy arrival. As such, nodes within a decoding set
$\mathbf{D}_i$ will generally have different time references. This
concept is illustrated in figure \ref{fig:time_reference_univ}.
Non-aligned time references obviously results in asynchronous
transmissions by relays. It is important at this point to study the
implications of asynchronous relaying on the delay spread at the
destination $Q$. For an arbitrary node $R_{i,k}$, the propagation
delay of the 1st energy arrival with respect to the packet source is
given by the following recursive formula
\begin{equation}
d_{p_{i,k}}=\min\{\overline{R_{i,k}R_{i-1,n}}+d_{p_{i-1,n}}+\delta_r\}_{n=1}^{\widetilde{K}_{i-1}}.\label{eq:prop_del}
\end{equation}
In (\ref{eq:prop_del}) we have expressed propagation delay in terms
of distance rather than time for notational convenience. The term
$\delta_r$ in Equation \ref{eq:prop_del} represents the difference
in length between the specular path and the first multipath echo of
the Rayleigh channel. We recall that Rayleigh channels do not
include a specular path. For the sake of simplicity, we assume that
$\delta_r$ is approximately equal for all pairs of nodes selected
from within two consecutive hops. The forwarding delay spread at the
destination $Q$ can then be expressed as
\begin{eqnarray}
\max\{\overline{QR_{q-1,k}}+d_{p_{q-1,k}}\}_{k=1}^{\widetilde{K}_{q-1}}-\min\{\overline{QR_{q-1,k}}+d_{p_{q-1,k}}\}_{k=1}^{\widetilde{K}_{q-1}}.
\label{eq:del_spread_Q}
\end{eqnarray}
The total delay spread is naturally the sum of the forwarding delay
spread and the multipath delay spread. At the first glance, the
discussion above may induce the impression that the forwarding delay
spread may grow indefinitely as packets progress towards the
destination. However, a closer look at the issue suggests otherwise.
The first few relays to receive and then transmit the packet at hop
$i$ are typically those who are the \textit{closest} to the
$(i-1)$th hop. At the same time, they are typically the
\textit{farthest} from the $(i+1)$th hop and thus will have the
largest propagation delays. It is straightforward to validate this
intuition analytically for a linear network. In fact, it can be
shown that for a linear network, packet copies are aligned in time
every other hop, i.e. $i$ is even. This is true when $B$ is
sufficiently large such that we may ignore the probability of having
negative progress nodes. For 2-D networks however, it is quite more
involved to validate such an intuition analytically. Rather it can
be more conveniently verified through simulations. figure
\ref{fig:delay_spread} illustrates the mean and standard deviation
of the forwarding delay spread for strip widths between 100 and 200
meters. Simulations have been carried out for 3 different values of
node density, $\rho$. As can be inferred from the figure, the mean
and standard deviation are almost independent of the node density.
It is also shown that for this range of strip width, the mean is
approximately 2$\mu$s with a standard deviation of no more than
0.35$\mu$s. This is valuable information as it provides guidance on
the suitable length of the cyclic prefix. For instance the length of
the cyclic prefix in LTE is at least 4.7$\mu$s while it is 10$\mu$s
for the IEEE 802.16e standard. Hence it is possible to utilize any
of these two standardized OFDM radios for OMR. This is not the case
for the IEEE 802.11g standards where the duration of the the cyclic
prefix is only 0.8$\mu$s. It can be further observed from figure
\ref{fig:del_spr_hist} that the forwarding delay spread tends to
follow a normal distribution. Moreover, figure
\ref{fig:delay_spread_mean_std} indicates that the standard
deviation of the forwarding delay spread increases with the strip
width, which is quite intuitive.

\subsection{Effect of Time Offset on RACH Signals}
\begin{figure}[t]
\begin{center}
\epsfxsize=9cm \centerline{\epsffile{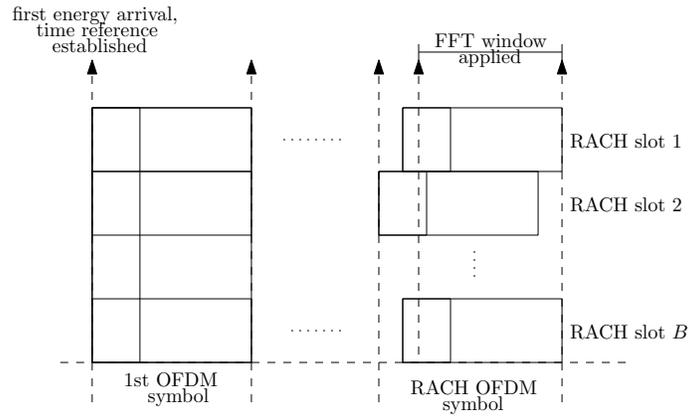}}
\end{center}
\caption{For some RACH slots the FFT window will not be aligned to
the actual start of the RACH signal.} \label{fig:rach}
\end{figure}

The signal of a non-empty RACH slot is composed of the superposition
of the location information of one or more relays. Each RACH slot is
randomly picked by a unique set of relays. As such, each RACH signal
undergoes a different channel towards a given receiver. Therefore,
RACH signals are generally expected to be non-aligned in time, as
exemplified in figure \ref{fig:rach}. Figure \ref{fig:rach} is only
showing first energy arrivals, i.e. subsequent signal echoes are not
depicted. The RACH OFDM symbol is the aggregation of all the RACH
signals. As shown in the figure, the receiver aligns its time
reference to the first energy arrival of the first OFDM symbol. The
FFT window is applied every integer multiple of the symbol duration.
As a result, some RACH signals will be suffering from a time offset
with respect to the start of the FFT window. The effect of time
offset on the detection of OFDM symbols was studied in detail in
\cite{ofdm.sync.offset}. It was shown in \cite{ofdm.sync.offset}
that when the time offset is ``towards'' the CP, i.e. the FFT window
is partially applied on the CP, then only a phase error is
introduced. Interestingly, it is only time offsets towards the CP
are possible in the case of OMR. If differential PSK is adopted,
then the effect of the time offset is greatly marginalized.

\section{Statistical Modeling of Hopping
Dynamics}\label{sec:hopping_dyn} 
\begin{figure}[t]
\begin{center}
\epsfxsize=9cm
\centerline{\epsffile{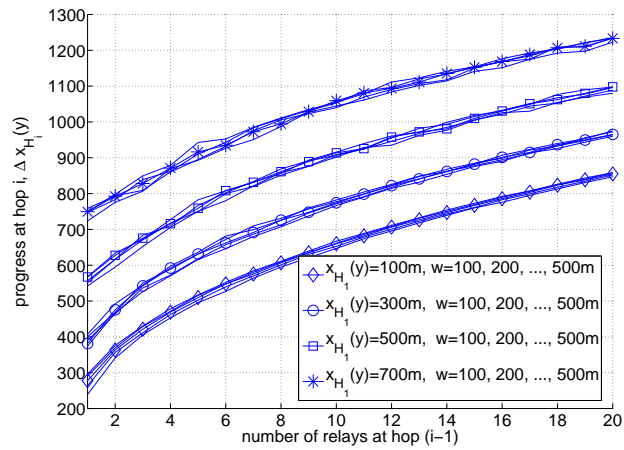}}
\end{center}
\caption{Simulation results demonstrating that the progress made
every hop is linear in $\widetilde{K}_{i-1}$. The impact of the
progress made in the $1$st hop, $x_{H_1}(y)$, carries on to the
subsequent hops. In other words, the larger $x_{H_1}(y)$ is the
greater progress is achieved in subsequent hops.}
\label{fig:linear_in_K}
\end{figure}

\begin{figure}[t]
\begin{center}
\epsfxsize=9cm \centerline{\epsffile{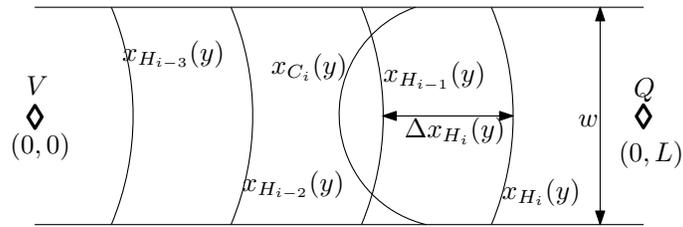}}
\end{center}
\caption{Setup to study hopping behavior of OMR.}
\label{fig:hop_dyn_setup}
\end{figure}

In this section, an elaborate
statistical framework is constructed for the sake of capturing the
dynamics of packet hopping under OMR. We start off by recalling that
given $\widetilde{K}_i$ relays in $\mathbf{R}_i$, then $j_i$
represents the index of the $1$st relay in $\mathbf{R}_i$ whose
position information is resolvable. Our next goal is to derive an
expression for the probability density function (PDF)
$p_{j_i}(j_i|\widetilde{K}_i)$. To proceed, we consider that the
ordered set $\mathbf{R}_i$ can be expressed as a block of length
$\widetilde{K}_i$ constructed from the alphabet $\{0,1\}$. Here
``$0$'' represents the event of being unresolvable while ``$1$''
represents the complementary event. Furthermore, we represent the
``do not care'' state with ``$\mathrm{x}$''. Accordingly,
$p_{j_i}(j_i|\widetilde{K}_i)$ is equivalent to
$\mathbb{P}[R_{i,1}R_{i,2}\ldots R_{i,{j_{i}}}\ldots
R_{i,\widetilde{K}_i}=\underbrace{00\ldots 01}_{j_i}
\mathrm{x}\ldots \mathrm{x}]$. To proceed, it is more convenient
first to derive an expression for the probability
$\mathbb{P}[R_{i,1}R_{i,2}\ldots R_{i,{j_{i}}}\ldots
R_{i,\widetilde{K}_i}=\underbrace{00\ldots 0}_{j_i-1}
\mathrm{x}\ldots \mathrm{x}]$ which equals
$1-(\mathbb{P}[\underbrace{00\ldots 01}_{j_i-1} \mathrm{x}\ldots
\mathrm{x}]+\mathbb{P}[\underbrace{00\ldots 10}_{j_i-1}
\mathrm{x}\ldots \mathrm{x}]+\ldots+\mathbb{P}[\underbrace{11\ldots
11}_{j_i-1} \mathrm{x}\ldots \mathrm{x}])$. Let us define
$\mathfrak{B}^{(m)}(\cdot)$ as a decimal-to-binary conversion
operator and $\mathfrak{C}^{(n,m)}(\cdot)$ as a cyclic shift right
operator, where $m$ corresponds to the size of the binary word and
$n$ is the order of the shift operator. It can be shown that the set
represented by $\{\mathfrak{B}^{(j_i-1)}(k)\}_{k=1}^{2^{j_i-1}-1}$
is completely covered by
\begin{eqnarray}
\{\mathfrak{C}^{(n,j_i-1)}(1\mathrm{x}\ldots
\mathrm{x})\}_{n=0}^{j_i-2} -
\{\mathfrak{C}^{(n,j_i-1)}(11\mathrm{x}\ldots
\mathrm{x})\}_{n=0}^{j_i-3} + \ldots
(-1)^{j_i-1}\{\mathfrak{C}^{(0,j_i-1)}(11\ldots
1)\}\label{eq:covered}.
\end{eqnarray}
Furthermore, we define $p_z$ as the probability of having exactly
$z$ relays in $\mathbf{R}_i$ whose position information are
resolvable. $p_z$ can be evaluated recursively such that
\begin{eqnarray}
p_z = \left\{
\begin{array}{ll}
\left(\frac{B-1}{B}\right)^{\widetilde{K}_i-1} & ,z=1\\
p_{z-1}\left(\frac{B-z}{B-z+1}\right)^{\widetilde{K}_i-z} &
,z=2\ldots B-2\\
0 & , z\geq B-1
\end{array}
\right.\label{eq:pz}
\end{eqnarray}
Consequently, it follows from (\ref{eq:covered}) and (\ref{eq:pz})
that
\begin{eqnarray}
\mathbb{P}[R_{i,1}R_{i,2}\ldots R_{i,{j_{i}}}\ldots
R_{i,\widetilde{K}_i}=\underbrace{00\ldots 0}_{j_i-1}
\mathrm{x}\ldots \mathrm{x}]=1+\sum_{z=1}^{j_i-1}(-1)^z {j_i-1
\choose z}p_z.
\end{eqnarray}
Now since $\mathbb{P}[R_{i,1}R_{i,2}\ldots R_{i,{j_{i}}}\ldots
R_{i,\widetilde{K}_i}=\underbrace{00\ldots 01}_{j_i}
\mathrm{x}\ldots \mathrm{x}]$ is equal to
$\mathbb{P}[R_{i,j_i}=1]\mathbb{P}[R_{i,1}\ldots
R_{i,{j_{i}-1}}=00\ldots 0|R_{i,j_i}=1]$, then we obtain
\begin{eqnarray}
p_{j_i}(j_i|\widetilde{K}_i)=\left\{
\begin{array}{ll}
\left(\frac{B-1}{B}\right)^{\widetilde{K}_i-1}\left(1+\sum_{z=1}^{j_i-1}(-1)^z
{j_i-1 \choose z}p_z\right) & ,j_i=1\ldots \widetilde{K}_i\\
1+\sum_{z=1}^{\widetilde{K}_i}(-1)^z {j_i-1 \choose z}p_z & ,j_i=0
\end{array}
\right.\label{eq:pj}
\end{eqnarray}
where $j_i=0$ refers to the event of all relays being unresolvable.
The probabilities $p_z$ and $p_1$ in (\ref{eq:pj}) are re-evaluated
by replacing $B$ in (\ref{eq:pz}) with $B-1$. Since
$p_1>\ldots>p_{z-1}>p_z$, it can be shown that
$\frac{dp_{j_i}(j_i|\widetilde{K}_i)}{dB}<0$, i.e.
$p_{j_i}(j_i|\widetilde{K}_i)$ decreases monotonically in $B$. This
is intuitive since the number of RACH collisions is expected to
decrease as $B$ increases.\\
\indent Another important modeling aspect is to characterize the
mean coverage contour at each hop. From (\ref{eq:xHi}), it can be
shown that $x_{H_i}(y), |y|\leq \pm \frac{w}{2}$ is concave, i.e.
$x_{H_i}(y)$ has a single maximum in $|y|\leq \pm \frac{w}{2}$. This
indicates that $x_{H_i}(y)$ may be approximated by a circular arc.
Furthermore, as $w$ increases, so does $\widetilde{K}_{i-1}$ (on
average). For two strip widths, $w_1$ and $w_2$, where $w_2>w_1$, we
have $x_{H_i}(y,w_2)>x_{H_i}(y,w_1)$. This suggests that if
$x_{H_i}(y)$ is to be approximated by an arc, then its radius
depends on $w$ and thus can be expressed as $\Omega w^c$, where
$\Omega$ and $c$ are network-dependent constants. Indeed, the
circularity of the coverage contour and the dependency of its radius
on $w$ have been validated numerically through a sufficient number
of simulations. Those simulations have also revealed that the
progress made every hop may be approximated by a linear function in
$\widetilde{K}_{i-1}$ as demonstrated in figure
\ref{fig:linear_in_K}. In other words, $\Delta
x_{H_i}(y)=x_{H_i}(y)-x_{H_{i-1}}(y)$ is equivalent to $\varphi
\widetilde{K}_{i-1}+\beta x_{H_1}(0)$, where $\varphi$ and $\beta$
are network-dependent constants, and
$x_{H_1}(0)=1/U^{\frac{1}{\alpha}}$. Consequently, we get the
following recursive relationship
\begin{eqnarray}
x_{H_i}(y)=x_{H_{i-1}}(y)+\varphi
\widetilde{K}_{i-1}+\frac{\beta}{U^{\frac{1}{\alpha}}}=\varphi\sum_{n=1}^{i-1}\widetilde{K}_n+(i-1)
\frac{\beta}{U^{\frac{1}{\alpha}}}+x_{H_1}(y).\label{eq:xH}
\end{eqnarray}
The intuition behind the linear approximation in (\ref{eq:xH}) may
be better perceived by considering the hypothetical case of having
$\widetilde{K}_{i-1}$ co-located relays. In such a case, we have
$x_{H_i}(y)=x_k+\sqrt{(\widetilde{K}_{i-1}/U)^{2/\alpha}-(y-y_k)^2}$
using (\ref{eq:xHi}). It can be shown that a linear approximation
here is good enough to provide a mean absolute percentage error
(MAPE) of 5.5$\%$ (and as low as 3$\%$ for $\widetilde{K}_{i-1}>3$).\\
\indent To further study the hopping behavior of OMR, we consider
the problem setup shown in figure \ref{fig:hop_dyn_setup}. For
narrow strips ($w/L$ relatively small), the decision contour
$x_{C_{i}}(y)$ can be conveniently assumed to be axially centered
around the line $y=0$. We recall that for hop $i-1$, we denoted the
$1$st relay in $\mathbf{R}_{i-1}$ whose position information is
resolvable by $R_{{i-1},j_{i-1}}$. We need to find an expression for
the distance of $R_{i-1,j_{i-1}}$ away from the point
$(x_{H_{i-1}}(0),0)$ which is in return approximately equal to the
distance $x_{H_{i-1}}(0)-x_{C_{i}}(0)$. The expectation of the
distance to the $n$th nearest neighbor in a sector with angle $\phi$
was derived in \cite{routing.fading.Haenggi} and is given by
$\sqrt{\frac{2}{\phi \epsilon\rho}(n-1-\frac{\pi}{4})}$. With
reference to figure \ref{fig:hop_dyn_setup}, the sector angle $\phi$
can be approximated to extend from $(x_{H_{i-1}}(0),0)$ to the
intercepts of $x_{H_{i-1}}(y)$ with $|y|=\frac{w}{2}$. Thus, $\phi$
can be estimated to be quite close to $\pi$. Therefore we get
\begin{eqnarray}
x_{C_i}(0)\approx x_{H_{i-1}}(0)- \sqrt{\frac{2}{\pi \epsilon
\rho}\left(j_{i-1}-1-\frac{\pi}{4}\right)}.\label{eq:xC}
\end{eqnarray}
We further define the following
areas (where $|y|\leq \pm \frac{w}{2}$):
\begin{eqnarray}
\mathcal{A}_{D_i}&=&\between\{x_{H_{i-1}}(y),x_{H_{i}}(y)\}\label{eq:A_D_i},\\
\mathcal{A}_{R_i}&=&\between\{x_{C_{i}}(y),x_{H_{i-1}}(y),x_{H_{i}}(y)\}\label{eq:A_R_i},\\
\mathcal{A}_{D_i}^-&=&\between\{x_{H_{i-3}},x_{H_{i-1}}\}\label{eq:A_Di-},\\
\mathcal{A}_{R_i}^-&=&\between\{x_{C_{i}}(y),x_{H_{i-1}}(y)\}\label{eq:A_Ri-}.
\end{eqnarray}
Moreover, the number of relays able to decode the packet at its
$i$th hop is denoted by $\widetilde{L}_i=|\mathbf{D}_i|$. In fact,
$\widetilde{L}_i$ is composed of two terms such that
$\widetilde{L}_i=L_i+L_i^{-}$. The terms $L_i$ and $L_i^{-}$
correspond to nodes lying in $\mathcal{A}_{D_i}$ and
$\mathcal{A}_{D_i}^-$ respectively. We note that $L_i^{-}$ can be
further broken down into two components. The first corresponds to
the nodes in $\between\{x_{H_{i-2}},x_{H_{i-1}}\}$ who were asleep
at the time when $\mathbf{R}_{i-2}$ began their transmission but
woke up before the transmission ended. Whereas the second
encompasses the nodes in $\between\{x_{H_{i-3}},x_{H_{i-2}}\}$ who
were asleep during the whole transmission duration of
$\mathbf{R}_{i-3}$ and only woke up before the transmission of
$\mathbf{R}_{i-2}$ ended. For simplification and conciseness of the
subsequent analysis, we assume that the sleeping time is equal to
$T_p$ such that the second component will have a value of zero and
thus (\ref{eq:A_Di-}) reduces to
$\mathcal{A}_{D_i}^-=\between\{x_{H_{i-2}},x_{H_{i-1}}\}$.
Similarly, we can define $\widetilde{K}_i=|\mathbf{R}_i|$ as
$\widetilde{K}_i=K_i+K_i^{-}$, where $K_i$ and $K_i^{-}$ are the
relays lying in $\mathcal{A}_{R_i}$ and $\mathcal{A}_{R_i}^-$
respectively. $K_i^{-}$ represents the nodes lying in
$\mathcal{A}_{R_i}^-$ who were asleep at the time when
$\mathbf{R}_{i-2}$ began their transmission but woke up before the
transmission ended.\\
\indent In order to evaluate the energy and delay performance, we
need to evaluate the expectations $\mathbb{E}[\widetilde{L}_i]$ and
$\mathbb{E}[\widetilde{K}_i]$. A suitable starting point is to study
the statistical dependencies of the areas $\mathcal{A}_{R_i}$,
$\mathcal{A}_{D_i}$, $\mathcal{A}_{D_i}^-$, and
$\mathcal{A}_{R_i}^-$. Based on (\ref{eq:xH}) and in light of
(\ref{eq:A_D_i}), it is clear the statistics of $\mathcal{A}_{D_i}$
are completely encompassed by $\widetilde{K}_{i-1}$. Similarly, from
(\ref{eq:xH}), (\ref{eq:xC}), and (\ref{eq:A_R_i}) it can be shown
that the statistics of $\mathcal{A}_{R_i}$ are dictated by
$j_{i-1}$, $\{\widetilde{K}_n\}_{n=1}^{i-2}$, and
$\{\widetilde{K}_n\}_{n=1}^{i-1}$. On the other hand, using
(\ref{eq:xH}) and (\ref{eq:A_Di-}), and recalling the assumption
that the sleep time is equal to $T_p$ it is evident that the
statistics $\mathcal{A}_{D_i}^-$ can be equivalently represented by
those of $\widetilde{K}_{i-2}$. Finally, since $\mathcal{A}_{R_i}^-$
is confined by $x_{C_{i}}(y)$ and $x_{H_{i-1}}(y)$ then using
(\ref{eq:xH}), (\ref{eq:xC}), and (\ref{eq:A_Ri-}) the statistics of
$\mathcal{A}_{R_i}^-$ are actually encompassed by $j_{i-1}$ and
$\{\widetilde{K}_n\}_{n=1}^{i-2}$. Given
$S_{i-v}=\sum_{n=1}^{i-v}\widetilde{K}_n$ and based on the
statistical dependencies exposed above, we subsequently obtain
\begin{eqnarray}
p_{L_i}(L_i)&=&\sum_{\widetilde{K}_{i-1}}
p_{L_i}(L_i|\mathcal{A}_{D_i})p_{\widetilde{K}_{i-1}}(\widetilde{K}_{i-1}),\label{eq:pLi}\\
p_{K_i}(K_i)&=&\sum_{j_{i-1}}\sum_{S_{i-1}}\sum_{S_{i-2}}
p_{K_i}(K_i|\mathcal{A}_{R_i})p(j_{i-1},S_{i-1},S_{i-2}),\label{eq:pKi}\\
p_{L_i^-}(L_i^-)&=&\sum_{\widetilde{K}_{i-2}}
p_{L_i^-}(L_i^-|\mathcal{A}_{D_i}^-)p_{\widetilde{K}_{i-2}}(\widetilde{K}_{i-2}),\label{eq:pLim}\\
p_{K_i^{-}}(K_i^{-})&=&\sum_{j_{i-1}}\sum_{S_{i-2}}
p_{K_i^{-}}(K_i^{-}|\mathcal{A}_{R_i}^-)p(j_{i-1},S_{i-2})\label{eq:pKim}.
\end{eqnarray}
To take one step ahead, the probability that a node in
$\mathcal{A}_{D_i}^-$ or similarly in $\mathcal{A}_{R_i}^-$ was
sleeping when $\mathbf{R}_{i-2}$ started transmitting is
$p_{wk}=1-\epsilon$. Recalling that nodes are dispersed in the field
according to a 2-D poisson distribution, we obtain
\begin{eqnarray}
p_{L_i}(L_i|\mathcal{A}_{D_i})&=&\frac{1}{L_i!}(\epsilon \rho
\mathcal{A}_{D_i})^{L_i}e^{-\epsilon \rho \mathcal{A}_{D_i}},\label{eq:possionLi}\\
p_{K_i}(K_i|\mathcal{A}_{R_i})&=&\frac{1}{K_i!}(\epsilon \rho
\mathcal{A}_{R_i})^{K_i}e^{-\epsilon \rho \mathcal{A}_{R_i}},\label{eq:possionKi}\\
p_{L_i^{-}}(L_i^{-}|\mathcal{A}_{D_i}^-)&=&\frac{1}{L_i^{-}!}(\epsilon
\rho \mathcal{A}_{D_i}^- p_{wk})^{L_i^{-}}e^{-\epsilon \rho
\mathcal{A}_{D_i}^- p_{wk}},\label{eq:possionLim}\\
p_{K_i^{-}}(K_i^{-}|\mathcal{A}_{R_i}^-)&=&\frac{1}{K_i^{-}!}(\epsilon
\rho \mathcal{A}_{R_i}^- p_{wk})^{K_i^{-}}e^{-\epsilon \rho
\mathcal{A}_{R_i}^- p_{wk}}.\label{eq:possionKim}
\end{eqnarray}
We note that $p_{S_{i-2}}(S_{i-2})=
\ast_{n=2}^{i-2}\left(p_{S_{n-1}}(S_{n-1}),p_{\widetilde{K}_n}(\widetilde{K}_n)\right)$,
where $\ast$ is the convolution operator. The joint PDF
$p(j_{i-1},S_{i-2})$ is given by $\sum_{S_{i-1}=0}^\infty
p(j_{i-1},S_{i-1},S_{i-2})$, where $p(j_{i-1},S_{i-1},S_{i-2})$
equals the product of $p_{j_{i-1}}(j_{i-1}|S_{i-1},S_{i-2})$,
$p_{S_{i-1}}(S_{i-1}|S_{i-2})$, and $p_{S_{i-2}}(S_{i-2})$.
Furthermore, $p_{S_{i-1}}(S_{i-1}|S_{i-2})$ is nothing but
$p_{\widetilde{K}_{i-1}}(S_{i-1}-S_{i-2})$. Also,
$p_{j_{i-1}}(j_{i-1}|S_{i-1},S_{i-2})$ is equivalent to
$p_{j_{i-1}}(j_{i-1}|S_{i-1}-S_{i-2})$ and can be computed by
evaluating (\ref{eq:pj}) at $i-1$ instead of $i$ and substituting
$\widetilde{K}_{i-1}$ with $S_{i-1}-S_{i-2}$ . With
$p_{K_i^{-}}(K_i^{-}|\mathcal{A}_{R_i}^-)$ and $p(j_{i-1},S_{i-2})$
readily available, we are now able to compute $p_{K_i^{-}}(K_i^{-})$
from (\ref{eq:A_Ri-}), (\ref{eq:pKim}), and (\ref{eq:possionKim}) .
On the flip side of the coin, $p_{K_i}(K_i)$ can be computed from
(\ref{eq:A_R_i}), (\ref{eq:pKi}), and (\ref{eq:possionKi}) knowing
that the joint PDF $p(j_{i-1},S_{i-1},S_{i-2})$ has already been
computed above. Since $\widetilde{K}_i=K_i+K_i^{-}$, then
$p_{\widetilde{K}_i}(\widetilde{K}_i)=p_{K_i}(K_i)\ast
p_{K_i^{-}}(K_i^{-})$. From the statistical analysis presented thus
far, it becomes clear that the statistics at any arbitrary hop can
be conveniently obtained by recursion. In other words, at hop $i$,
$p_{\widetilde{K}_{i-1}}(\widetilde{K}_{i-1})$ and
$p_{\widetilde{K}_{i-2}}(\widetilde{K}_{i-2})$ will be readily
available. Accordingly, $p_{L_i}(L_i)$ is obtained from
(\ref{eq:A_D_i}), (\ref{eq:pLi}), and (\ref{eq:possionLi}) while
$p_{L_i^-}(L_i^-)$ is obtained from (\ref{eq:A_Di-}),
(\ref{eq:pLim}), and (\ref{eq:possionLim}).\\
\indent Finally, we are in a position now to derive an expression
for the expected number of retransmissions $\mathbb{E}[n_{r_i}]$
occurring at hop $i$. Given the areas $\mathcal{A}_{R_i}$ and
$\mathcal{A}_{R_i}^-$, then
$\mathbb{E}[n_{r_i}|\mathcal{A}_{R_i},\mathcal{A}_{R_i}^-]=1/(e^{\epsilon
\rho (\mathcal{A}_{R_i}+p_{wk}\mathcal{A}_{R_i}^-)}-1)$. If the
number of RACH slots $B$ is sufficiently large, then the
contribution of $\mathcal{A}_{R_i}^-$ diminishes and may be
overlooked in this expression for convenience and tractability of
the analysis. Hence, we get:
\begin{equation}
\mathbb{E}[n_{r_i}]=\sum_{j_{i-1}}\sum_{S_{i-1}}\sum_{S_{i-2}}
\mathbb{E}[n_{r_i}|\mathcal{A}_{R_i}]p(j_{i-1},S_{i-1},S_{i-2}).\label{eq:nri}
\end{equation}

\section{Performance Evaluation}\label{sec:eval}
To appreciate the end-to-end performance of OMR against other
beaconless protocols, we are going to evaluate it in light of the
inherent tradeoff between energy and delay. Denoting the mean
end-to-end energy consumed in forwarding one packet by $E_{e2e}$ and
the mean end-to-end delay by $l_{e2e}$, we define the end-to-end
energy-delay product as $\mathrm{EDP}=E_{e2e}l_{e2e}$. In order to
account for the fact that OMR and beaconless protocols may employ
different modulation and coding schemes (MCS), $\mathrm{EDP}$ must
be normalized by the PHY data rate, $r$. Hence, we can define an
end-to-end cost metric $C_{e2e}=\frac{\mathrm{EDP}}{rT_p}$ which
reflects the amount of energy consumed and delay time spent in
transporting $rT_p$ bits from source to destination.

\subsection{OMR Performance Metrics}
At any given hop $i$, there would be $\widetilde{K}_{i-1}$ nodes who
are relaying the packet and $\widetilde{L}_i$ nodes receiving the
transmission. At the next hop, $i+1$, the $\widetilde{K}_{i-1}$
relays would be listening to make sure the packet is being forwarded
ahead. Consequently,
\begin{equation}
l_{e2e}=T_p\sum_{i=1}^q (1+\mathbb{E}[n_{r_i}]),
\end{equation}
where $q:x_{H_q}(0)\geq L$ is the number of hops traversed by the
packet to the destination. Furthermore, the energy expended at hop
$i$ to relay the packet ahead is given by:
\begin{eqnarray}
E_i&=&\mathbb{E}[\widetilde{L}_{i}]P_{Rx}T_p+(\mathbb{E}[n_{r_i}]+1)\mathbb{E}[\widetilde{K}_{i-1}]P_tT_p+\nonumber\\
&&(\mathbb{E}[n_{r_i}]+1)\mathbb{E}[\widetilde{K}_{i-1}]P_{Rx}t_{ID}+(\mathbb{E}[\widetilde{K}_{i-1}]t_{ID}+\mathbb{E}[\widetilde{L}_{i}]T_p)\frac{P_t}{N_s},
\label{eq:Ei}
\end{eqnarray}
where $P_{Rx}$ is the the power consumed in receiving a packet and
$\frac{P_t}{N_s}$ is the busy tone power. The end-to-end energy
consumed is $E_{e2e}=\sum_{i=1}^q E_i$.

\subsection{A Spotlight on the Performance of Beaconless Protocols}
The family of beaconless position-based forwarding protocols
includes quite a few variants. Nevertheless, the work of
\cite{GeRaF.latency.zorzi,GeRaF.multihop.zorzi} constituted a major
stepping stone towards the development of other beaconless
protocols. In addition, the analytical framework provided in
\cite{GeRaF.latency.zorzi,GeRaF.multihop.zorzi} is quite
comprehensive and detailed. As such, it is going to be adopted in
this paper as the benchmark for comparison. Other beaconless
protocols may be considered to a great extent as adaptations and/or
enhancements of \cite{GeRaF.latency.zorzi,GeRaF.multihop.zorzi}.
Thus, evaluation results of this section can be conveniently
generalized to other beaconless protocols. For the sake of brevity,
we will be often using the term ``BCL'' to refer to the class of
existing beaconless position-based protocols. Table
\ref{table:beaconless} explains the different stages of the BCL
forwarding process. At a given hop, there would be $\eta$ empty
cycles followed by one non-empty cycle. Empty cycles occur when
there are no awaken nodes offering positive progress within the
transmission range of the sender. The transmission range of the
sender is denoted by $d_m$. On average, the fraction of nodes within
the transmission range offering positive progress towards the
destination is $\xi$. Each cycle consists of $N_p$ slots such that
the duration of one slots is $T_s$. For the non-empty cycle, there
would be $m_e$ empty slots followed by $m_n$ collision-resolution
slots. Summing up all terms of energy consumption, the mean energy
consumed in transmitting one packet at a certain hop is given by
\begin{eqnarray}
P_tT_s((\mathbb{E}[m_e]+5+\mathbb{E}[\eta]
N_p)N_s+(1+2\mathbb{E}[m_e]\xi)\epsilon\rho\pi d_{m}^2
+1\nonumber\\
+ \mathbb{E}[m_e]+ \mathbb{E}[\eta] N_p + \xi N_s \epsilon\rho\pi
d_{m}^2 / N_p + (2+3N_s)(\mathbb{E}[m_n]-1))/N_s.
\end{eqnarray}
On the other hand, the mean energy consumed in receiving:
\begin{eqnarray}
P_{Rx}T_s((1+2\xi \mathbb{E}[m_e])\epsilon\rho\pi d_{max}^2 + 2
+\mathbb{E}[m_e]+\mathbb{E}[\eta] N_p + 3(\mathbb{E}[m_n]-1)).
\end{eqnarray}
The expectations $\mathbb{E}[\eta]$, $\mathbb{E}[m_e]$, and
$\mathbb{E}[m_n]$ are found in explicit forms in
\cite{GeRaF.latency.zorzi} ((3) and (4)). The end-to-end delay is
function of the time spent per hop and the number of hops traversed
before reaching the destination. In light of table
\ref{table:beaconless}, and \cite{GeRaF.latency.zorzi} ((4), (5),
and (16)), it can be shown that the time spent on average by a
packet in the case of average beaconless protocols is given by
$2\left(\mathbb{E}[\eta]N_p+\mathbb{E}[m_e+m_n]\right)T_s$.
Furthermore, the expected progress towards the destination after $i$
hops is given by \cite{GeRaF.multihop.zorzi}, ((8) and (19)).

\subsection{OMR vs. BCL}
\begin{figure}[t]
\begin{center}
\epsfxsize=10cm \centerline{\epsffile{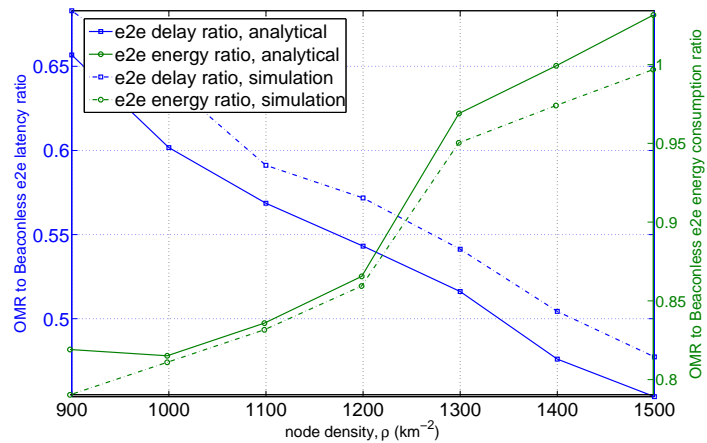}}
\end{center}
\caption{End-to-end energy and delay performance of OMR compared to
beaconless protocols. OMR is utilizing here a transmit power that is
9 dB less than beaconless.} \label{fig:hop_dyn}
\end{figure}

OMR performance was evaluated from an analytical point of view in
light of the framework provided in Section \ref{sec:hopping_dyn}. It
was then compared to
\cite{GeRaF.latency.zorzi,GeRaF.multihop.zorzi}. Furthermore,
simulations have been carried out to validate the outcomes of the
analytical computations. The strip width $w$ was set to $200$ m in
the simulations with a source-destination separation of $2$ km. The
sleeping duty cycle was set to $\epsilon=25\%$ while the detection
threshold $\gamma_t$ was assumed to be $5$ dB at $\tau=20\%$. The
path loss exponent was considered
to be $\alpha=3$.\\
\indent The main theme to be conveyed in this section, is that OMR
starts to strikingly outperform existing beaconless position-based
protocols in terms of end-to-end delay as node density grows.
However, this comes at the price of additional energy consumption
since a larger number of nodes tend to relay the packet. This
reasserts the significance of evaluating end-to-end performance
based on the interaction between energy and delay. We are going to
show in this section that there is more than one scenario whereby
OMR outperforms BCL protocols from the joint perspective of delay
and energy. For instance, it will be demonstrated that by tuning
down the transmit power of OMR, delay performance is not
substantially impacted while energy consumption is noticeably
reduced. Thus, OMR is able to provide an obvious performance gain in
this case. Moreover, OMR will be also shown to have an edge for
certain classes of modulation techniques.\\
Finally and before delving into the detailed comparison, it ought to
be mentioned that the outcome of analytical computations have been
closely matched by simulation results, as per figure
\ref{fig:hop_dyn}.

\subsubsection{Effect of Transmit Power}
\begin{figure}[t]
\begin{center}
\subfigure[End-to-end cost comparison as function of OMR transmit
power at various node densities.]{
\includegraphics[width=10cm]{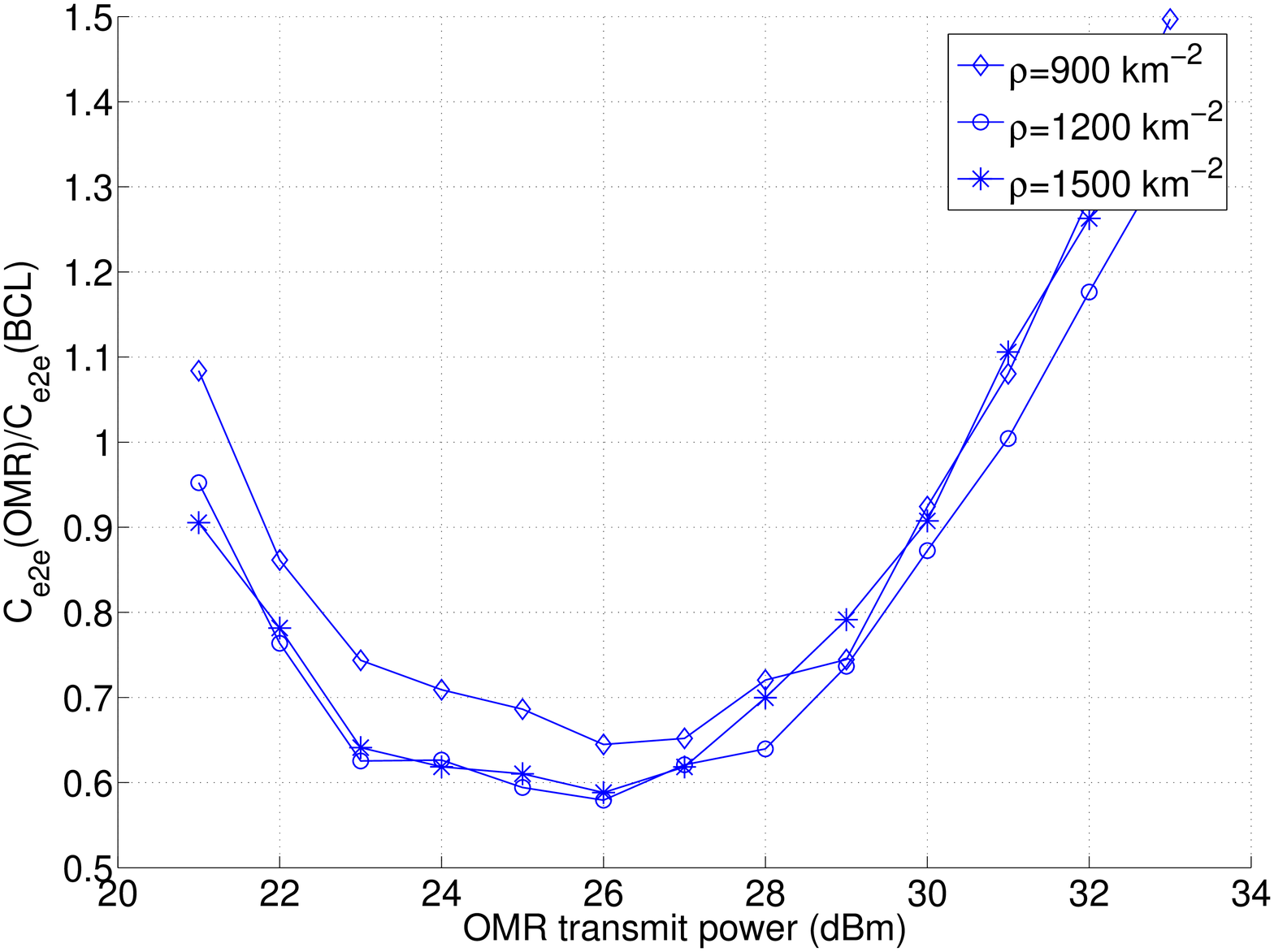}\label{fig:EDP_power2}}
\subfigure[End-to-end energy-delay product of OMR divided by that of
the beaconless protocol. The duration of the control packet (RTS,
CTS, ...) utilized by the beaconless protocol has been varied to
study its impact on the performance. Here, OMR is operating at a
transmit power which is 9 dB less than that of the beaconless.]{
\includegraphics[width=10cm]{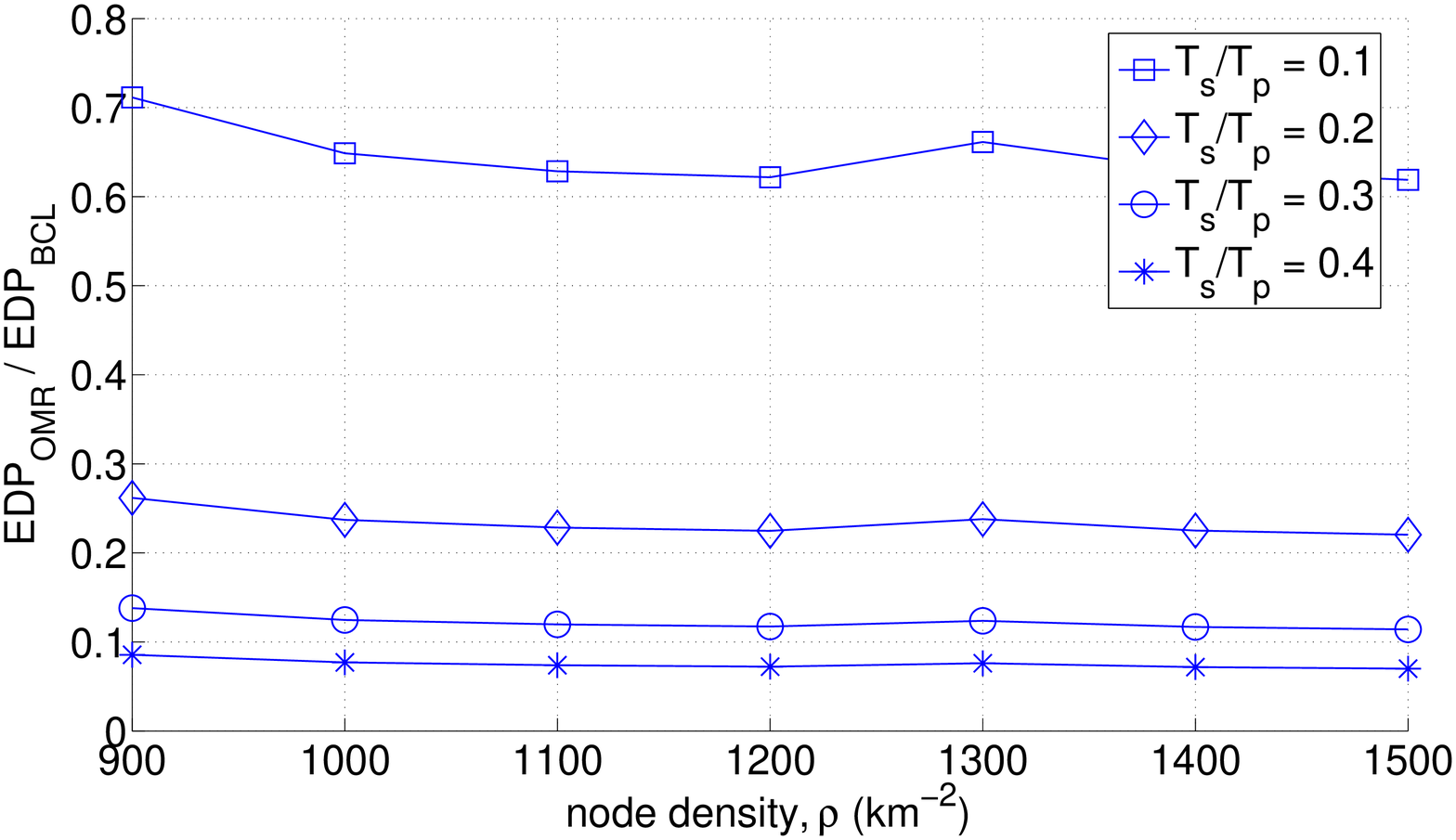}\label{fig:EDP_Ts}}
\end{center}
\caption{Comparing performance of OMR from and end-to-end
perspective to beaconless protocols.}
\label{fig:performance_comparison_1}
\end{figure}

It is clearly demonstrated from figure \ref{fig:hop_dyn} that for an
equivalent amount of energy consumption, OMR offers reduced
end-to-end delay. This conclusion is mainly valid when OMR utilizes
a lower transmit power. We have reverted to using a lower transmit
power for OMR based on the rationale that it reduces energy
consumption significantly while not really jeopardizing the hop
distances traversed by packets. This argument stems from the simple
fact that the relationship between the transmit power and the
achievable hop distance is inversely scaled by the path loss
exponent $\alpha$. As such, a substantial drop in transmit power
will be countered by only moderate to marginal shrink in hop
distance, depending on the value of $\alpha$. Indeed, this has been
verified by plotting the ratio
$C_{e2e}(\mathrm{OMR})/C_{e2e}(\mathrm{BCL})$ against OMR's transmit
power for various node densities. Results are depicted in figure
\ref{fig:EDP_power2} which illustrates that OMR offers roughly a
$40\%$ enhancement when the transmit power is in the range of 6 to 9
dB below that of BCL (which is operating here at 33 dBm).
Nevertheless, reducing the transmit power further below a certain
threshold will actually start to have a counter effect. As the
transmit power decreases, the probability of retransmission starts
to pick up quickly and rather contributes to the inflation of OMR's
$\mathrm{EDP}$. It can be also concluded from figure
\ref{fig:EDP_power2} that the performance gains offered by OMR is
almost indifferent of the underlying node density. Indeed, this is a
design objective which has been set forth early on in this paper.\\
\indent Since OMR does not resort to any collision resolution
mechanism, it is also of interest to investigate the effect of the
RTS/CTS packet duration, denoted by $T_s$, relative to the data
packet duration, $T_p$. As $T_s/T_p$ increases, a considerable
portion of the end-to-end energy consumption in the beaconless case
is attributed to RTS/CTS transmissions during the collision
resolution process. End-to-end delay as well increases. Figure
\ref{fig:EDP_Ts} illustrates that for shorter packet lengths, or
alternatively larger $T_s/T_p$ ratios, OMR is set to offer
substantial performance gains
over beaconless protocols.\\
\subsubsection{Effect of The Number of RACH Slots, $B$} For the sake of a comprehensive and fair
comparison, we have also studied the impact of $B$ on the
performance of OMR. On one hand, increasing $B$ will reduce the
probability of RACH collisions. This in return will have the effect
of reducing the size of the area $\mathcal{A}_{R_i}^-$ and thus will
result in reducing energy consumption. The delay performance will
not be affected noticeably since the size of $\mathcal{A}_{R_i}^-$
is typically small compared to $\mathcal{A}_{R_i}$. On the other
hand, as $B$ gets larger, the overhead at the PHY layer also grows
effectively resulting in a reduction of the effective data rate seen
at layer 2. This intuition is indeed validated in figure
\ref{fig:cost_e2e_B}.
\subsubsection{Higher Order MCS} 
\begin{figure}[t]
\begin{center}
\subfigure[End-to-end cost of OMR at various values of $B$ (number
of RACH slots) divided by that of the beaconless protocol.]{
\includegraphics[width=10cm]{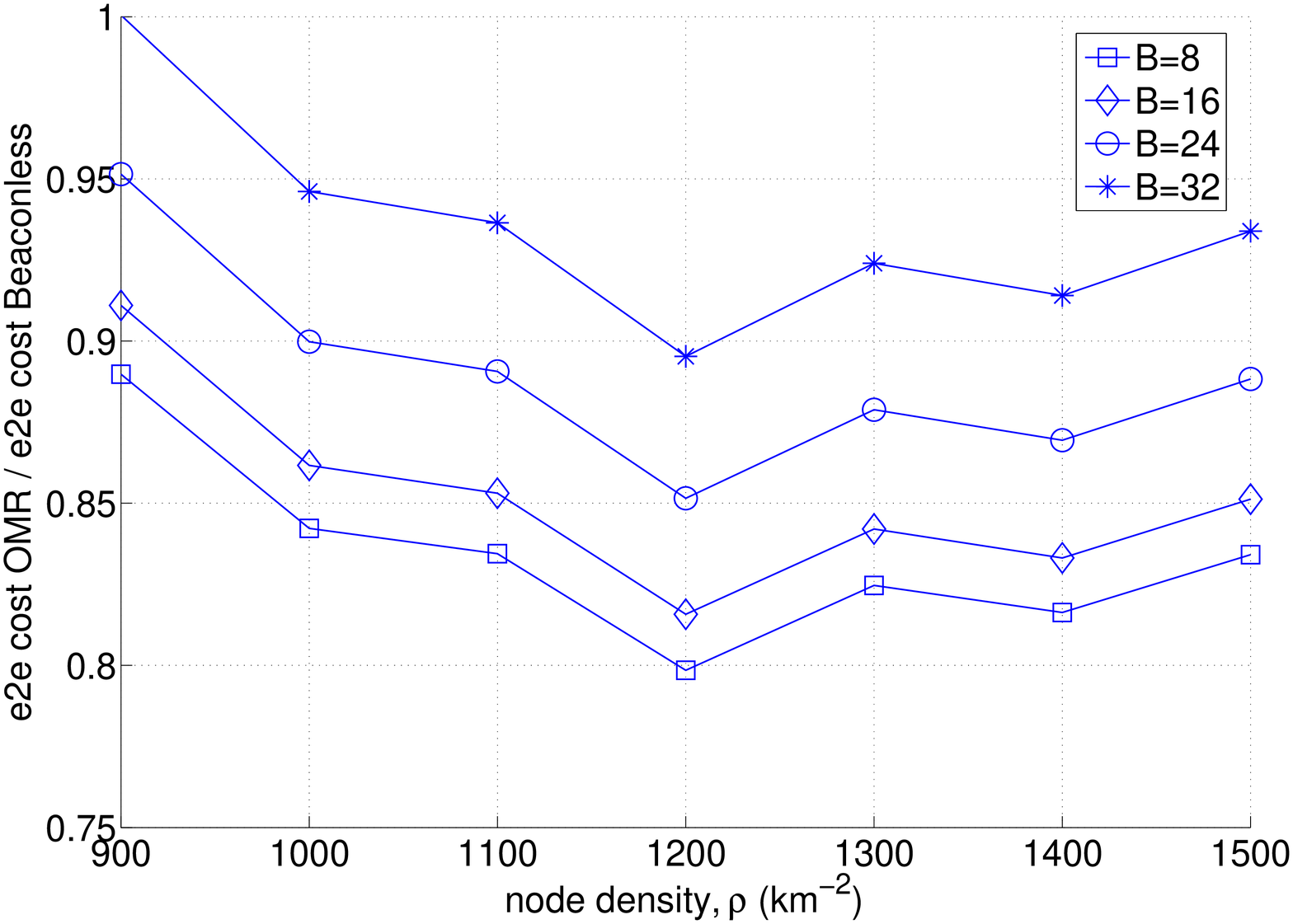}\label{fig:cost_e2e_B}}
\subfigure[End-to-end cost of OMR at various modulation schemes
divided by that of the beaconless protocol.]{
\includegraphics[width=10cm]{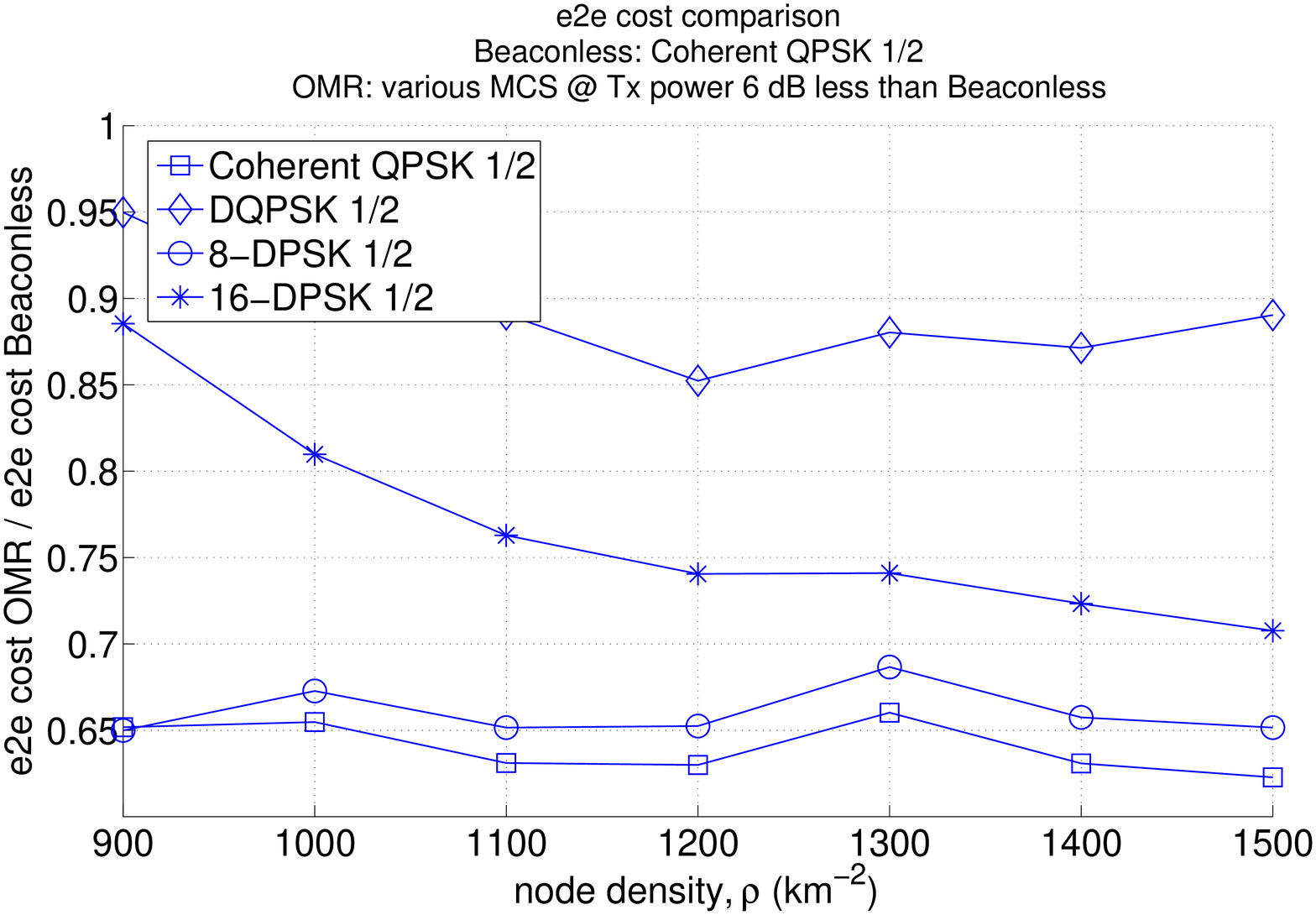}\label{fig:cost_e2e_MCS}}
\end{center}
\caption{Comparing performance of OMR from and end-to-end
perspective to beaconless protocols, continued.}
\label{fig:performance_comparison_2}
\end{figure}

\begin{figure}[t]
\begin{center}
\subfigure[Snapshots of two concurrent packet relaying processes]{
\includegraphics[width=15cm]{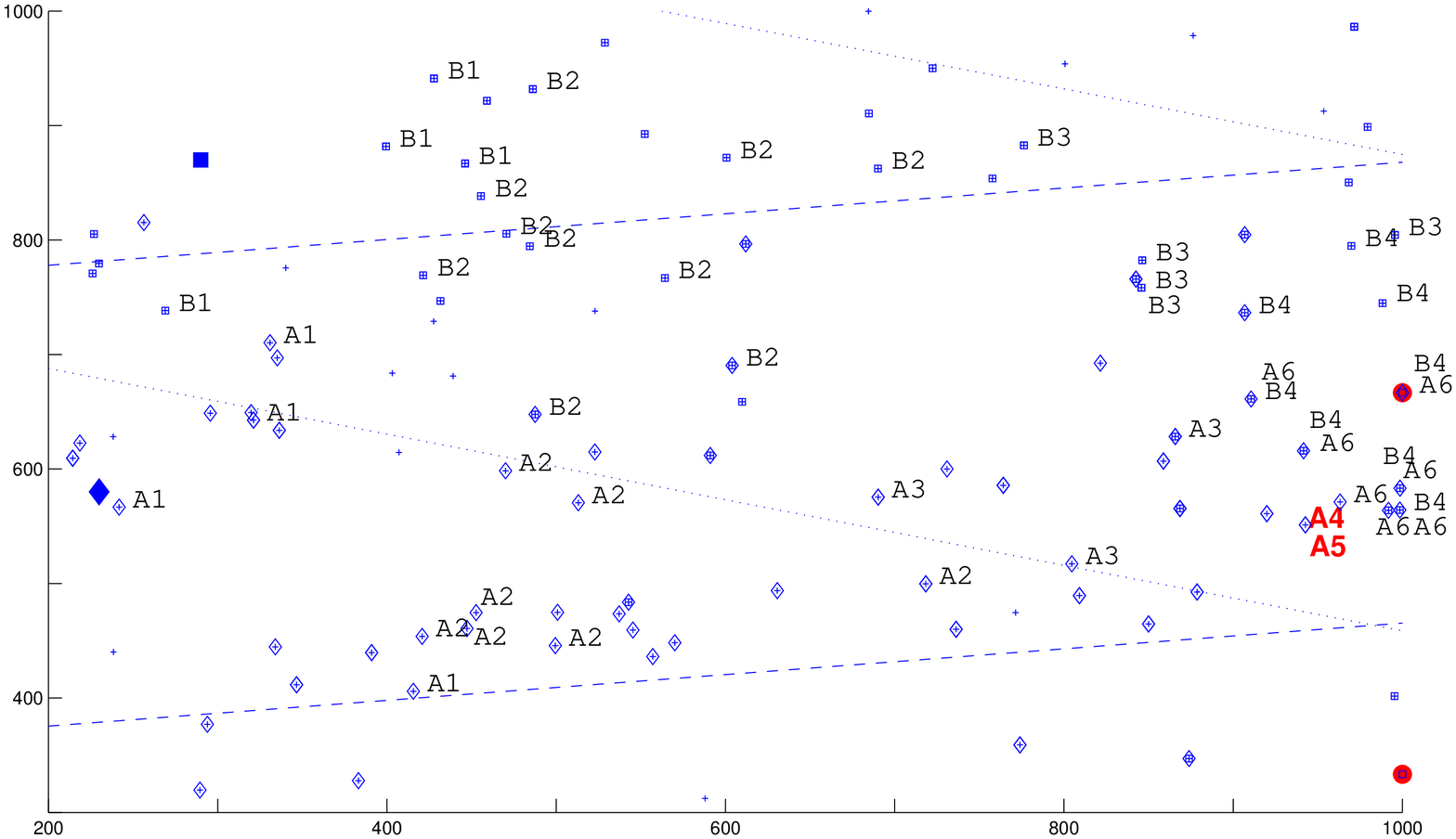}\label{fig:sample_process}}
\subfigure[Legend]{
\includegraphics[width=9cm]{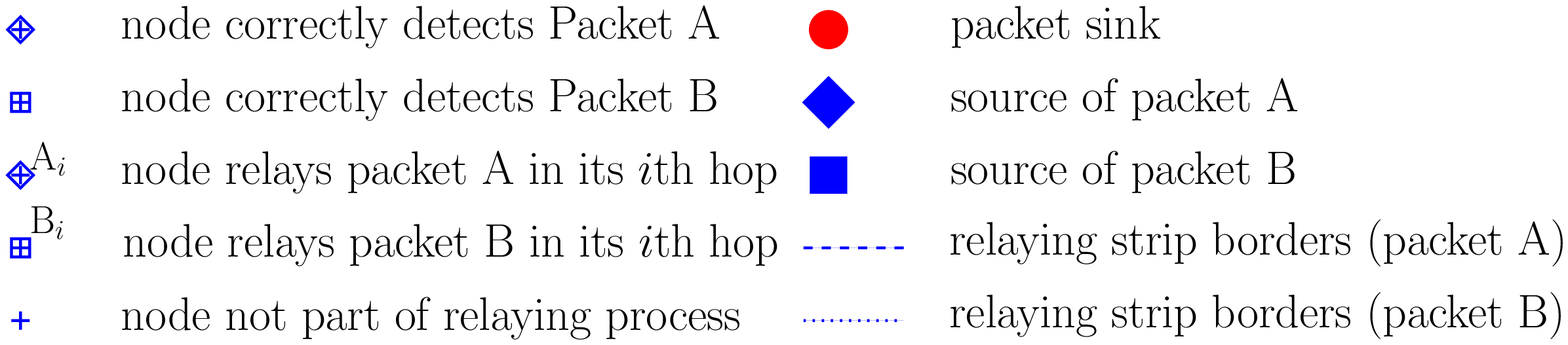}\label{fig:legend}}
\end{center}
\caption{Demonstrating the flow of two concurrent packets in the
network under OMR.} \label{fig:sample relaying process}
\end{figure}

The performance gains demonstrated thus far are actually
encouraging to consider a higher order MCS for OMR. We recall here
our original choice to deploy a differential modulation scheme in
conjunction with OMR. Whereas coherent modulation necessitates
accurate estimation of the channel fading coefficients, differential
modulation does not; thus reducing cost and complexity of the
receiver. The tradeoff however in using M-DPSK instead of coherent
M-PSK is a higher detection threshold. Now, our next task would be
to specify the detection threshold $\gamma_t$ for each MCS under
consideration for OMR. The bit error rate (BER) targeted here is
$10^{-2}$. Under the assumption of two-branch maximal ratio
combining (MRC) and Gray encoding, then the detection thresholds for
DQPSK, 8-DPSK, and 16-DPSK are approximately 12.8 dB, 15.6, and 18.5
dB respectively \cite{fadingbook}. There is no need to consider
2-DPSK as it has the same detection threshold of DQPSK. On the other
hand, for coherent QSPK, the required detection threshold at BER of
$10^{-2}$ approximately evaluates to 10.85 dB \cite{fadingbook}.
Furthermore, if rate $\frac{1}{2}$ convolutional coding with a
constraint length of 2 is employed then a coding gain of $3$ to $4$
dB can be achieved (at a BER of $10^{-2}$)
\cite{coding.gain.Yasuda}. Figure \ref{fig:cost_e2e_MCS} depicts the
end-to-end cost incurred by OMR in comparison to BCL for various
MCSs. BCL is assumed to use coherent QPSK. Figure
\ref{fig:cost_e2e_MCS} clearly conveys that OMR is able to transport
the same amount of bits much faster while consuming the same amount
of energy. Looking at it from a complementary angle, OMR delivers
more bits towards the destination (by utilizing a higher order MCS)
while consuming the same amount of energy and spending the same
amount of delay.\\
\indent Finally, simulations have been also carried out to
investigate the behavior of OMR in case of forwarding two concurrent
but dissimilar packets. A sample forwarding process is depicted in
figure \ref{fig:sample relaying process} noting that both packets
have the same destination. Interference between the two forwarding
processes has been accounted for in the simulation. It can be
observed that one retransmission has occurred at the 4th hop of
Packet A. This was due to the significant interference induced by
the forwarding process of Packet B. As Packet B progressed further
towards the destination, the forwarding process of Packet A gained
some room to resume.

\section{Conclusions}\label{sec:conclusions}
In this paper we have proposed a novel multi-relay beaconless
position-based packet forwarding protocol for WSNs. The protocol
couples the use of an OFDM-based PHY with position-based routing to
create an improved end-to-end performance over traditional
beaconless protocols. A statistical framework has been provided to
study the hopping dynamics and behavior of the proposed protocol.
Numerical and simulation results have shown that the proposed
protocol may be tuned to offer an improvement of up to 40$\%$ in
terms of the end-to-end performance.\\
\indent In our future work, we will study the applicability and
performance of OMR in case of using the hop count away from the
destination instead of geographical positions. We will also analyze
the co-channel interference created by one forwarding process on
another, evaluate how it limits the overall network capacity, and
research means to control it.

\balance
\bibliographystyle{IEEEbib11}
\bibliography{bibfileshort}


\begin{table}[t]
\caption{Typical energy consumption components for a beaconless
protocol.}
\begin{center}
\begin{tabular}{|llll|}
  \hline\hline

  \multicolumn{4}{|c|}{\textbf{$\eta$ empty cycles}}\\

  \hline\hline

  \textbf{Node(s)} & \textbf{Count} & \textbf{Activity} & \textbf{Duration}\\

  sender & 1 & transmit RTS & $T_s$\\

         & &listening and activating BT while listening & $\eta N_p
         T_ s$\\

         &  &transmits CONTINUE message after each slot not containing CTS & $\eta N_p T_ s$\\

  \hline\hline

  \multicolumn{4}{|c|}{\textbf{non-empty cycle}}\\

  \hline\hline

  \textbf{Node(s)} & \textbf{Average Count} & \textbf{Activity} & \textbf{Duration}\\

  sender & 1 & transmit RTS & $T_s$\\


  sender & 1 & listening and activating BT while listening &
  $m_eT_s$\\
  & &transmits CONTINUE message after each slot not containing CTS & $m_e T_ s$\\

  potential relays & $\frac{(N_p-m_e)}{N_p}\xi\epsilon\rho\pi d_{m}^2$ & listen to the
  channel in anticipation of a CTS message & $m_eT_s$\\
  & & listen to CONTINUE messages from sender & $m_eT_s$\\
  & & activate BT while listening & $2m_eT_s$\\

  timers expired & $\frac{\xi\epsilon\rho\pi d_{m}^2}{N_p}$ &
  transmit CTS message in $(m_e+1)$th slot & $T_s$\\

  colliding & at least 2 & transmit CTS, listen for CTS-Reply from
  sender, activate BT while listening & $(m_n-1)T_s$\\

  sender & 1 & transmit CONTINUE message & $(m_n-1)T_s$\\
          &   & receive colliding CTS messages & $(m_n-1)T_s$\\

  successful relay & 1 & transmit CTS message during $(m_e+m_n)$th slot & $T_s$\\

  sender & 1 & listen to CTS from successful relay during $(m_e+m_n)$th
  slot & $T_s$\\
        &    & activate BT while listening  & $T_s$\\
        &    & transmit OK message  & $T_s$\\

   successful relay & 1 & listen to OK message and activate BT while listening & $T_s$\\
   \hline\hline

\end{tabular}
\end{center}
 \label{table:beaconless}
\end{table}

\end{document}